\documentclass[11pt]{article}

\usepackage[paper=a4paper,dvips,top=2.0cm,left=2.2cm,right=2.2cm,bottom=2.9cm,footskip=1.6cm]{geometry}
\usepackage{amsfonts,amsmath,amsthm,amssymb}
\numberwithin{equation}{section}  
\usepackage[svgnames]{xcolor}
\usepackage{fix-cm}
\usepackage{sectsty}
\usepackage{fancyhdr}
\usepackage{lastpage}
\usepackage{graphicx}
\usepackage{url}
\usepackage{enumerate}
\usepackage{paralist}
\usepackage{multicol}
\usepackage{color}  
\usepackage{float}  
\usepackage{epsfig}
\usepackage{hyperref}   
\hypersetup{colorlinks=true,linkcolor=beamer@PRD, citecolor=beamer@PRD}
\usepackage{authblk}  
\usepackage{cite}  

\definecolor{beamer@blue}{RGB}{0,0,255}
\definecolor{beamer@mediumblue}{RGB}{0,0,190}
\definecolor{beamer@midnightblue}{RGB}{25,25,112}
\definecolor{beamer@navy}{RGB}{0,0,128}
\definecolor{beamer@darkblue}{RGB}{0,0,139}
\definecolor{beamer@purple}{RGB}{128,0,128}
\definecolor{beamer@levander}{RGB}{100.,149.,237.}
\definecolor{beamer@PRD}{RGB}{46,48,146}
\definecolor{beamer@green}{RGB}{0,128,0}
\definecolor{beamer@darkgreen}{RGB}{0,100,0}
\definecolor{beamer@olive}{RGB}{128,128,0}
\definecolor{beamer@darkolivegreen}{RGB}{85,107,47}
\definecolor{beamer@gray}{RGB}{190,190,190}
\definecolor{beamer@ivry}{RGB}{220,220,220}
\definecolor{beamer@new}{RGB}{40,120,50}
\definecolor{shadecolor}{RGB}{220,220,220}
\definecolor{beamer@darkslategray}{RGB}{47,79,79}
\definecolor{beamer@chocolate}{RGB}{210,105,30}
\definecolor{beamer@brown}{RGB}{165,42,42}
\definecolor{beamer@orangered}{RGB}{255,69,0}
\definecolor{beamer@maroon}{RGB}{128,0,0}
\definecolor{beamer@white}{RGB}{234,242,243}
\definecolor{beamer@silver}{RGB}{0.5,0.45,0.37}

\begin{document}


\title{\textbf{Nonclassicality versus entanglement in a noncommutative space}}


\author{\textbf{Sanjib Dey$^{1,4}$, Andreas Fring$^{2}$ and V{\'e}ronique Hussin$^{1,3}$}}
\date{\footnotesize{$^{1}$Centre de Recherches Math{\'e}matiques (CRM), Universit{\'e} de Montr{\'e}al, Montr{\'e}al H3C 3J7, Qu{\'e}bec, Canada \\ $^{2}$Department of Mathematics, City University London, Northampton Square, London EC1V 0HB, UK \\ $^{3}$Department de Math{\'e}matiques et de Statistique, Universit{\'e} de Montr{\'e}al, Montr{\'e}al H3C 3J7, Qu{\'e}bec, Canada \\ $^{4}$Department of Mathematics and Statistics, Concordia University, Montr{\'e}al H3G 1M8, Qu{\'e}bec, Canada \\ \small{E-mail: dey@crm.umontreal.ca, a.fring@city.ac.uk, veronique.hussin@umontreal.ca}}}
\maketitle
    	
\thispagestyle{fancy}
\begin{abstract}
In a setting of noncommutative space with minimal length we confirm the general assertion that the more nonclassical an input state for a beam splitter is, the more entangled its output state becomes. By analysing various nonclassical properties we find that the odd Schr\"odinger cat states are more nonclassical than the even Schr\"odinger cat states, hence producing more entanglement, which in turn are more nonclassical than coherent states. Both the nonclassicality and the entanglement can be enhanced by increasing the noncommutativity of the underlying space.  In addition we find as a by-product some rare explicit minimum uncertainty quadrature and number squeezed states, i.e. ideal squeezed states. 
\end{abstract}	 
\begin{section}{Introduction} \label{sec1}
\addtolength{\footskip}{-0.4cm} 
\addtolength{\voffset}{1.2cm} 
Nonclassicality is a special property of light which can not be characterized by the classical theory of electromagnetism. The proper definition, rather, lies at the core of quantum optics. The density matrix $\hat{\rho}$ of any state of light can be expressed in terms of a coherent state $\vert\alpha\rangle$, $\hat{\rho}=\textstyle\int P(\alpha)\vert\alpha\rangle\langle\alpha\vert d^2\alpha$. For a classical state of light, $P(\alpha)$ represents a probability density function and, therefore, one expects $P(\alpha)\geq 0$, which is indeed true for coherent states. In fact, for coherent states the $P$ function is a delta function. However, for some quantum mechanical states, the $P$ function can be negative or more singular than a delta function, such that $P(\alpha)$ can no longer be interpreted as probability density function. In this sense, the states which correspond to negative $P$ functions or more singular than delta functions, are usually referred to \textit{nonclassical states} \cite{Glauber2,Sudarshan2}.

Squeezed states are the best examples of nonclassical states \cite{Walls,Loudon_Knight}. A state is \textit{quadrature squeezed}, if any of its quadratures has a standard deviation that falls below that of the coherent states. On the other hand, a \textit{number squeezed} state has photon number uncertainty lower than the coherent state. One of the exciting features of quadrature and number squeezed states is that they exhibit much more reduced noise than even a single photon state or a coherent state \cite{Walls,Loudon_Knight}. That is the reason why squeezed light; or more generally, nonclassical light has applications in optical communication \cite{Yamamoto_Haus}, optical measurement \cite{Caves}, gravitational wave detection \cite{Vahlbruch_et} and in many other scenarios. They are also utilised to distribute secret keys to perform quantum cryptography. Moreover, nonclassical states produce entangled states, which are the most fundamental requirement for quantum teleportation.

Throughout the history of quantum information there have been various attempts to investigate how nonclassical a quantum system is. One of the common techniques is to study different squeezing properties of the corresponding system. The more squeezed a quantum system is in its quadrature components and/or in photon number distribution, the more nonclassical it is. Quantum entanglement is another excellent feature in particular, which quantifies the nonclassicality of the state independently \cite{Kim_Son_Buzek_Knight}. In this paper, we study both of the approaches; i.e. we first analyse various squeezing properties of our systems in a complete analytical fashion and, thereafter, we compute the entanglement of the corresponding systems to verify our analytical findings. For the purpose of analysis we have chosen the famous Schr\"odinger cat states \cite{Schrodinger_cat} and coherent states in a noncommutative structure arising from the generalised uncertainty relation \cite{Kempf_Mangano_Mann,Bagchi_Fring,Dey_Fring_Gouba}. However, our principal findings are the advantages of utilising the noncommutative models instead of the usual quantum mechanical models; for instance, both nonclassicality and entanglement are being enhanced when we switch on the noncommutative parameter. By switching to noncommutative space, we also find an existence of both quadrature and number squeezed state satisfying the equality in the uncertainty relation; i.e. the \textit{ideal squeezed state}, which is very rare in the literature.    

Our manuscript is organised as follows: In Sec. \ref{sec2}, we discuss the general construction procedure of nonlinear coherent and cat states. In Sec. \ref{sec3}, we assemble various generalities from Sec. \ref{sec2} and construct the coherent and cat states for a specific harmonic oscillator in a noncommutative space. In Sec. \ref{sec4}, several nonclassical properties of noncommutative coherent states have been explored analytically. The analytical results are then verified independently by computing the entanglement of the corresponding nonclassical states.  In Sec. \ref{sec5}, we perform a similar task as Sec. \ref{sec4}, however, for noncommutative cat states instead. Our conclusions are stated in Sec. \ref{sec6}. 
\end{section}
\begin{section}{Nonlinear coherent and cat states}\label{sec2}
Coherent states play an important role in various branches of physics including mathematical physics, quantum gravity \cite{Freidel_Livine}, quantum cosmology \cite{Robles_Hassouni_Gonzalez},  quantum optics \cite{Gerry_Knight_Book}, atomic and molecular physics \cite{Gazeau_Book} etc. Their behaviour very closely resemble classical particles \cite{Dey_Fring_PRA}. The light fields which are prepared for the development of lasers, are very close to such states. Coherent states, that reduce the optical noise to a sufficiently low extent, can be obtained from any of the following simple definitions \cite{Glauber}: (i) as eigenstates of the annihilation operator $a\vert\alpha\rangle=\alpha\vert\alpha\rangle$, (ii) by applying the Glauber's unitary displacement operator, $D(\alpha)=\text{exp}(\alpha a^\dagger-\alpha^\ast a)$ on the vacuum state and (iii) as quantum states that minimize the uncertainty relation $\Delta x^2\Delta p^2=\hbar^2/4$, with equal uncertainties in each coordinate, $\Delta x^2=\Delta p^2$ \cite{Loudon_Knight}. However, every coherent state does not satisfy all of the above properties at a same time; for instance, see \cite{Antoine_Gazeau_Monceau_Klauder_Penson,Dey_Fring_Gouba_Castro}, where the first two properties have been satisfied, but not the third one. Coherent states that satisfy all of the three properties, are usually specified to \textit{intelligent states} \cite{Aragone_Guerri_Salamo_Tani}.
\lhead{Nonclassicality versus entanglement in a nc space}
\chead{}
\rhead{}

Generalizations of coherent states have been carried out since early 1970s \cite{Perelomov,Nieto_Simmons,Filho_Vogel,Manko_Marmo_Sudarshan_Zaccaria,Gazeau_Klauder} using different approaches resulting to different types of states. The most common technique follows from the replacement of the bosonic creation and annihilation operators $a,a^\dagger$ by the generalised ladder operators $A, A^\dagger$, such that
\begin{equation}\label{DefLadder}
A^\dagger = f(n)a^\dagger, \qquad A = af(n),
\end{equation}
where $f(n)$ is an operator-valued function of the number operator $n=a^\dagger a$. The deformed operators, $A$ and $A^\dagger$ obey the following nonlinear commutator algebras:
\begin{equation}\label{DefAlgeb}
[A,A^\dagger]=(n+1)f^2(n+1)-n f^2(n), \quad [n,A]=-A, \quad  [n,A^\dagger]=A^\dagger,
\end{equation} 
where the nonlinearity arises from $f(n)$. However, with the choice of $f(n)=1$, the deformed algebras \textcolor{beamer@PRD}{(}\ref{DefAlgeb}\textcolor{beamer@PRD}{)} reduce to the Heisenberg algebra. The eigenvalue definition of harmonic oscillator coherent states leads to the generalised version, which can be represented in number state basis as follows 
\begin{eqnarray}\label{NlCoherent}
\vert\alpha,f\rangle &=& \frac{1}{\tilde{\mathcal{N}}(\alpha,f)}\displaystyle\sum_{n=0}^{\infty}\frac{\alpha^n}{\sqrt{n!}f(n)!}\vert n\rangle, \qquad  \alpha\in \mathbb{C}~,
\end{eqnarray}
where $f(0)=1$ and the normalisation constant follows from the requirement $\langle\alpha,f\vert\alpha,f\rangle=1$,
\begin{eqnarray}
\tilde{\mathcal{N}}^2(\alpha,f) &=& \displaystyle\sum_{n=0}^{\infty}\frac{\vert\alpha\vert^{2n}}{n!f^2(n)!}~.
\end{eqnarray} 
The generalised and nonlinear coherent states have attracted a lot of interest in the literature for their classical like properties, for further details; see, \cite{Dey_Fring_squeezed,Ghosh_Roy,Dey_Fring_Gouba_Castro}. More fascinating states in quantum optics arise from the even and odd superposition of two coherent states, known as Schr\"odinger cat states \cite{Xia_Guo,Filho_Vogel_Cat,Mancini,Gerry_Knight,Sivakumar_Cat,Roy,Wu_Yang,Dey}, which are defined as follows 
\begin{eqnarray}\label{NlCat}
\vert\alpha,f\rangle_{\pm} &=& \frac{1}{\tilde{\mathcal{N}}(\alpha,f)_{\pm}}\Big(\vert\alpha,f\rangle \pm \vert -\alpha,f\rangle\Big),
\end{eqnarray} 
where the normalisation constant is 
\begin{eqnarray}
\tilde{\mathcal{N}}^2(\alpha,f)_{\pm} &=& 2\pm\frac{2}{\tilde{\mathcal{N}}^2(\alpha,f)}\displaystyle\sum_{n=0}^{\infty}\frac{(-1)^n\vert\alpha\vert^{2n}}{n!f^2(n)!}~.
\end{eqnarray} 
Sometimes, the states \textcolor{beamer@PRD}{(}\ref{NlCat}\textcolor{beamer@PRD}{)} are also familiar as even and odd coherent states. It is easy to check that the states \textcolor{beamer@PRD}{(}\ref{NlCat}\textcolor{beamer@PRD}{)} are eigenstates of the square of the generalised annihilation operator \textcolor{beamer@PRD}{(}\ref{DefLadder}\textcolor{beamer@PRD}{)}
\begin{equation}
A^2\vert\alpha,f\rangle_\pm =\alpha^2\vert\alpha,f\rangle_\pm~.
\end{equation}
The striking feature of cat states is that they are nonclassical in nature, which is very useful for creating  entangled states. Their behaviour is very similar to the squeezed states; such as, they acquire quadrature and amplitude (number) squeezing, photon bunching and anti-bunching etc \cite{Dey}. Sometimes the cat states are easier to generate than squeezed states as well. Moreover, cat states have been attempted in recent days to be  identified as basic states for logical qubit to generate quantum gates \cite{Gilchrist,Ourjoumtsev,Gao}, which makes them more interesting in the subject of quantum information processing.
\end{section} 
\begin{section}{Coherent and cat states for the perturbative noncommutative harmonic oscillator}\label{sec3}
The simplest and most commonly studied version of the noncommutative space-time structures consists of replacing the standard set of commutation relations for the canonical coordinates $x^\mu$ by the noncommutative versions, such as 
\begin{equation}
[x^\mu,x^\nu]=i\hbar \theta^{\mu\nu}, \quad [x^\mu,p_\nu]=i\hbar \delta^\mu_\nu \quad \text{and} \quad [p_\mu,p_\nu]=0,
\end{equation}
where $\theta^{\mu\nu}$ is a constant antisymmetric tensor. More fascinating structures, leading for instance to minimal lengths and generalised versions of Heisenberg's uncertainty relations, are obtained when $\theta^{\mu\nu}$ is taken to be functions of the momenta and coordinates. For further details on the physical implications of the latter version of the space-time structure, one may refer to \cite{Kempf_Mangano_Mann,Bagchi_Fring,Dey_Fring_Gouba,Dey_Fring_Khantoul,Dey_Fring_Time,Ali_Das_Vagenas,Nozari_Etemadi}. Here, we consider the one dimensional version of the algebra \cite{Bagchi_Fring,Dey_Fring_Gouba}
\begin{equation}\label{NcComm}
[X,P]=i\hbar(1+\check{\tau}P^2), \qquad X=(1+\check{\tau}p^2)x, \qquad P=p,
\end{equation} 
where $\check{\tau}=\tau/(m\omega\hbar)$ has the dimension of inverse squared momentum with $\tau$ being dimensionless. The noncommutative observables $X,P$ are represented in \textcolor{beamer@PRD}{(}\ref{NcComm}\textcolor{beamer@PRD}{)} in terms of the standard canonical variables $x,p$ satisfying $[x,p]=i\hbar$. The spectrum of the noncommutative harmonic oscillator
\begin{eqnarray}\label{NcHam}
H &=& \frac{P^2}{2m}+\frac{m\omega^2}{2}X^2-\hbar\omega \left(\frac{2+\tau}{4}\right),
\end{eqnarray} 
were computed by following the standard techniques of Rayleigh-Schr{\"o}dinger perturbation theory \cite{Dey_Fring_Gouba,Dey_Fring_squeezed} to the lowest order to
\begin{alignat}{1}
E_n &=\hbar\omega nf^2(n)=\hbar\omega n\Big[1+\frac{\tau}{2}(1+n)\Big]+\mathcal{O}(\tau^2), \\
\vert\phi_n\rangle &=\vert n\rangle-\frac{\tau}{16}\sqrt{(n-3)^{(4)}}\vert n-4\rangle+\frac{\tau}{16}\sqrt{(n+1)^{(4)}}\vert n+4\rangle+\mathcal{O}(\tau^2),\label{EigenF}
\end{alignat} 
where $Q^{(n)}:=\textstyle\prod_{k=0}^{n-1}(Q+k)$ denotes the Pochhammer symbol. The nonlinear coherent states \textcolor{beamer@PRD}{(}\ref{NlCoherent}\textcolor{beamer@PRD}{)} and cat states \textcolor{beamer@PRD}{(}\ref{NlCat}\textcolor{beamer@PRD}{)}, for the case at hand, are then defined as 
\begin{eqnarray}
\vert\alpha,f,\phi\rangle &=& \frac{1}{\mathcal{N}(\alpha,f)}\displaystyle\sum_{n=0}^{\infty}\frac{\alpha^n}{\sqrt{n!}f(n)!}\vert \phi_n\rangle, \qquad  \alpha\in \mathbb{C}~, \label{NcCoherent}\\
\vert\alpha,f,\phi\rangle_{\pm} &=& \frac{1}{\mathcal{N}(\alpha,f)_{\pm}}\Big(\vert\alpha,f,\phi\rangle \pm \vert -\alpha,f,\phi\rangle\Big), \label{NcCat}
\end{eqnarray}
where
\begin{alignat}{1}
f^2(n)!=\frac{\tau^n}{2^n}\left(2+\frac{2}{\tau}\right)^{(n)}, \quad \frac{1}{f^2(n)!} = 1-\frac{\tau}{4}n(3+n)+\mathcal{O}(\tau^2),
\end{alignat}
\begin{eqnarray}
\mathcal{N}^2(\alpha,f) &=& e^{\vert\alpha\vert^2}\left(1-\tau\vert\alpha\vert^2-\frac{\tau}{4}\vert\alpha\vert^4\right)+\mathcal{O}(\tau^2), \\
\mathcal{N}^2(\alpha,f)_\pm &=& 2\pm\frac{e^{-\vert\alpha\vert^2}}{2\mathcal{N}^2(\alpha,f)}\left(4+4\tau\vert\alpha\vert^2-\tau\vert\alpha\vert^4\right)+\mathcal{O}(\tau^2).
\end{eqnarray}
From now and onwards, we will drop the explicit mentioning of the order in $\tau$, understanding that all our computations are carried out to first order. Using the explicit form of $\vert\phi_n\rangle$ \textcolor{beamer@PRD}{(}\ref{EigenF}\textcolor{beamer@PRD}{)}, one can rewrite the coherent states \textcolor{beamer@PRD}{(}\ref{NcCoherent}\textcolor{beamer@PRD}{)} in the following form \cite{Dey_Hussin}
\begin{equation}\label{NcCoherentAlt}
\vert\alpha,f,\phi\rangle = \frac{1}{\mathcal{N}(\alpha,f)}\displaystyle\sum_{n=0}^{\infty}\frac{\mathcal{C}(\alpha,n)}{\sqrt{n!}f(n)!}\vert n\rangle,
\end{equation}
where
\begin{equation}
\mathcal{C}(\alpha,n)=\left\{ \begin{array}{lcl}
\alpha^n-\frac{\tau}{16}\alpha^{n+4}\frac{f(n)!}{f(n+4)!} & \mbox{if}
& 0\leq n \leq 3 \\ \alpha^n-\frac{\tau}{16}\alpha^{n+4}\frac{f(n)!}{f(n+4)!}+\frac{\tau}{16}\alpha^{n-4}\frac{n!}{(n-4)!}\frac{f(n)!}{f(n-4)!}\ & \mbox{if} & n\geq 4~.\end{array}\right.
\end{equation} 
The Hamiltonian \textcolor{beamer@PRD}{(}\ref{NcHam}\textcolor{beamer@PRD}{)} is clearly non-Hermitian with respect to the standard inner product. However, for a concrete realisation of physical reality of the Hamiltonian, one can follow the references \cite{Scholtz_Geyer_Hahne,Bender_Boettcher,Mostafazadeh_2002,Bender_Making_Sense} to construct the isospectral Hermitian counterpart of the non-Hermitian Hamiltonian \textcolor{beamer@PRD}{(}\ref{NcHam}\textcolor{beamer@PRD}{)}. For the construction, one considers the non-Hermitian Hamiltonian to be pseudo/quasi Hermitian; which means, the non-Hermitian Hamiltonian $H$ and the Hermitian Hamiltonian $\tilde{h}$ are related by the similarity transformation $\tilde{h}=\eta H \eta^{-1}$, with $\eta^\dagger\eta$ being a positive definite operator playing the role of the metric. Then, the corresponding eigenstates $\vert\Phi\rangle$ and $\vert\phi\rangle$ of $H$ and $\tilde{h}$, respectively, are related as $\vert\Phi\rangle=\eta^{-1}\vert\phi\rangle$. The same reasoning has to be adopted for the evaluation of the expectation values with regard to coherent states or cat states. Therefore, the expectation value for a non-Hermitian operator $\mathcal{O}$ related to a Hermitian operator $\tilde{o}$ by a similarity transformation $\tilde{o}=\eta \mathcal{O}\eta^{-1}$ is computed as
\begin{equation}\label{Similarity}
\langle\alpha,f,\Phi\vert\mathcal{O}\vert\alpha,f,\Phi\rangle_\eta :=\langle\alpha,f,\Phi\vert\eta^\dagger\eta\mathcal{O}\vert\alpha,f,\Phi\rangle =\langle\alpha,f,\phi\vert \tilde{o}\vert\alpha,f,\phi\rangle~.
\end{equation}
Our notation is to be understood in the sense that in the states $\vert\alpha,f,\Phi\rangle$ and $\vert\alpha,f,\phi\rangle$, we sum over the eigenstates of the non-Hermitian Hamiltonian $H$ and Hermitian Hamiltonian $\tilde{h}$, respectively. The Dyson map $\eta$, which relates the non-Hermitian Hamiltonian \textcolor{beamer@PRD}{(}\ref{NcHam}\textcolor{beamer@PRD}{)} to its isospectral Hermitian counterpart $\tilde{h}$, is found in this case to be $\eta=(1+\check{\tau}p^2)^{-1/2}$.  
\end{section}
\begin{section}{Noncommutative coherent state: an ideal squeezed state}\label{sec4}
We start our analysis on coherent states in noncommutative spaces by first evaluating the expectation values of the bosonic creation and annihilation operators
\begin{eqnarray}
\langle\alpha,f,\phi\vert a\vert\alpha,f,\phi\rangle &=& \alpha+\frac{\tau}{4}\Big(\alpha^{\ast3}-\alpha^2\alpha^\ast-2\alpha\Big), \\
\langle\alpha,f,\phi\vert a^\dagger\vert\alpha,f,\phi\rangle &=& \alpha^\ast+\frac{\tau}{4}\Big(\alpha^{3}-\alpha^{\ast 2}\alpha-2\alpha^\ast\Big).
\end{eqnarray} 
Let us now define two quadrature operators $y$ and $z$,
\begin{equation}\label{quadrature}
y=\frac{1}{\sqrt{2}}(a^\dagger+a), \qquad z=\frac{i}{\sqrt{2}}(a^\dagger-a),
\end{equation}
which are nothing but the dimensionless position and momentum operators $x,p$, respectively. Using the quadrature operators \textcolor{beamer@PRD}{(}\ref{quadrature}\textcolor{beamer@PRD}{)} and expanding $y^2$ and $z^2$ in terms of $a^\dagger$ and $a$, we compute the expectation values of each quadrature operators and their squares, which we do not present here. Instead, we report the more complicated computations for the noncommutative operators as follows
\begin{alignat}{1}
\langle\alpha,f,\Phi\vert Y\vert\alpha,f,\Phi\rangle_\eta &= \langle\alpha,f,\phi\vert \tilde{y}\vert\alpha,f,\phi\rangle = \frac{(\alpha+\alpha^\ast) \left[4-\tau  (\alpha-\alpha^\ast)^2\right]}{4 \sqrt{2}}, \label{Y}\\
\langle\alpha,f,\Phi\vert Z\vert\alpha,f,\Phi\rangle_\eta &= \langle\alpha,f,\phi\vert \tilde{z}\vert\alpha,f,\phi\rangle = \frac{i (\alpha -\alpha^\ast) \left[2\tau+\tau (\alpha +\alpha^\ast)^2-4\right]}{4 \sqrt{2}},
\end{alignat} 
where the noncommutative quadrature operators $Y,Z$ are mapped to the Hermitian operators $\tilde{y},\tilde{z}$ according to the similarity transformation $\tilde{y}=\eta Y\eta^{-1}=y+\tau(z^2y+yz^2)/2$ and $\tilde{z}=\eta Z\eta^{-1}=z$, as discussed before Eq. \textcolor{beamer@PRD}{(}\ref{Similarity}\textcolor{beamer@PRD}{)}. Using a similar procedure, albeit more lengthy computations yield
\begin{alignat}{1}
\langle\alpha,f,\Phi\vert Y^2\vert\alpha,f,\Phi\rangle_\eta &= \frac{1}{4}\Big[2+2(\alpha+\alpha^\ast)^2-\tau\big(\alpha^2+\alpha^{\ast 2}+\alpha^4+\alpha^{\ast 4}-4\vert\alpha\vert^2-2\vert\alpha\vert^4-2\big)\Big], \\
\langle\alpha,f,\Phi\vert Z^2\vert\alpha,f,\Phi\rangle_\eta &= \frac{1}{4}\Big[2-2(\alpha-\alpha^\ast)^2+\tau\big(\alpha^2+\alpha^{\ast 2}+\alpha^4+\alpha^{\ast 4}-4\vert\alpha\vert^2-2\vert\alpha\vert^4\big)\Big]. \label{Z2}
\end{alignat}
The classical like properties of coherent states in noncommutative spaces have been investigated thoroughly in an analytical fashion \cite{Dey_Fring_squeezed,Ghosh_Roy,Dey_Fring_Gouba_Castro}; albeit, for Gazeau-Klauder coherent states \cite{Gazeau_Klauder}. Here, we utilize the nonlinear coherent states and analyse the nonclassical properties instead, which are more useful for the purpose of quantum computation.
\begin{subsection}{Quadrature squeezing}\label{sec41}
Combining \textcolor{beamer@PRD}{(}\ref{Y}\textcolor{beamer@PRD}{)}-\textcolor{beamer@PRD}{(}\ref{Z2}\textcolor{beamer@PRD}{)}, we compute the variances of $Y$ and $Z$ as follows
\begin{equation}
\Delta Y^2 = R+\tau\left(\frac{1}{4}+\frac{\vert\alpha\vert^2}{2}\right), \quad \Delta Z^2 = R-\tau\left(\frac{1}{4}+\frac{\vert\alpha\vert^2}{2}\right).
\end{equation}
Whereas, the right hand side of the generalised uncertainty relation
\begin{equation}\label{GUR}
\Delta Y\Delta Z\geq\frac{1}{2}\Big\vert\langle\alpha,f,\Phi\vert [Y,Z]\vert\alpha,f,\Phi\rangle_\eta\Big\vert,
\end{equation}
is computed to
\begin{equation}\label{RHSGUR}
R=\frac{1}{2}\Big[1+\tau\langle\alpha,f,\Phi\vert Z^2\vert\alpha,f,\Phi\rangle_\eta\Big]=\frac{1}{4}\Big[2+\tau-\tau(\alpha-\alpha^\ast)^2\Big].
\end{equation}
It is easy to check that the generalised uncertainty relation \textcolor{beamer@PRD}{(}\ref{GUR}\textcolor{beamer@PRD}{)} is saturated in this case; i.e. $\Delta Y\Delta Z =R$ and, therefore, the coherent states in noncommutative space can be referred to intelligent states. However, unlike the coherent states of ordinary harmonic oscillator, uncertainties in two quadratures in this case are not equal to each other. Rather, the quadrature $Z$ is squeezed below the right hand side of the uncertainty relation $R$ \textcolor{beamer@PRD}{(}\ref{RHSGUR}\textcolor{beamer@PRD}{)}, whereas the quadrature $Y$ is expanded correspondingly, such that the uncertainty relation saturates. These types of nonclassical states; i.e. the intelligent states with squeezing in one quadrature, are very uncommon in the literature, which are known as the ``ideal squeezed states" \cite{Gerry_Knight_Book}. Similar type of states were found to exist in a slightly different context in \cite{Dey} by one of the authors, however, they seem to be less obvious to realize in the real life. Let us now investigate if the states are squeezed in photon number distribution as well.
\end{subsection}  
\begin{subsection}{Photon number squeezing}\label{sec42}
By photon number squeezing, it means the photon number distribution of the state being narrower than the average number of photons, i.e. $\langle(\Delta n)^2\rangle <\langle n\rangle$, with $n=a^\dagger a$ being the number operator. A simple way to recognise the photon statistics of any state is to calculate the Mandel parameter $Q$ \cite{Mandel},
\begin{eqnarray} \label{Mandel}
Q &=& \frac{\langle(\Delta n)^2\rangle}{\langle n\rangle}-1.
\end{eqnarray}
For states with $Q=0$, the statistics are Poissonian, while $Q>0$ and $Q<0$ correspond to the cases of super-Poissonian and sub-Poissonian statistics, respectively. For an ordinary coherent state, it is well known that the photon distribution is Poissonian $(Q=0)$ with a mean of $\langle n\rangle$. While, in noncommutative case we may follow Ref. \cite{Antoine_Gazeau_Monceau_Klauder_Penson} and expand the energy $E_n$ in the expression for the wave packet \textcolor{beamer@PRD}{(}\ref{NcCoherent}\textcolor{beamer@PRD}{)}
\begin{eqnarray}
\vert\alpha,f,\phi\rangle &=& \displaystyle\sum_{n=0}^\infty c_n(\alpha)\vert\phi_n\rangle,
\end{eqnarray}
with the weighting function $c_n(\alpha)=\alpha^n/\mathcal{N}(\alpha,f,\phi)\sqrt{\rho_n}$, about $\langle n\rangle=\frac{\vert\alpha\vert}{2}\frac{d}{d\vert\alpha\vert}\ln \mathcal{N}^2(\alpha,f,\phi)$. We then easily compute
\begin{eqnarray}
\langle N\rangle=\langle\alpha,f,\Phi\vert A^\dagger A\vert\alpha,f,\Phi\rangle_\eta &=& \vert\alpha\vert^2-\frac{\tau\vert\alpha\vert^2}{2}\big(2+\vert\alpha\vert^2\big), \\
\langle N^2\rangle=\langle\alpha,f,\Phi\vert A^\dagger AA^\dagger A\vert\alpha,f,\Phi\rangle_\eta &=& \vert\alpha\vert^2+\vert\alpha\vert^4-\tau \vert\alpha\vert^2\big(1+3 \vert\alpha\vert^2+\vert\alpha\vert^4\big),
\end{eqnarray}
such that
\begin{equation}
(\Delta N)^2 = \langle N^2\rangle-\langle N\rangle^2 = \vert\alpha\vert^2-\tau \big(\vert\alpha\vert^2+\vert\alpha\vert^4\big).
\end{equation}
As a consequence, the Mandel parameter \textcolor{beamer@PRD}{(}\ref{Mandel}\textcolor{beamer@PRD}{)} turns out to be negative, $Q=-\tau\vert\alpha\vert^2/2$, suggesting a sub-Poissonian statistics.  It is worthwhile to  mention that a state exhibiting sub-Poissonian statistics; $(Q<0)$, is said to possess the number squeezing and shows nonclassical behaviour \cite{Gerry_Knight_Book}. Therefore, the state that we are discussing here is very unique. It is ideal squeezed state as analysed in Sec. \ref{sec42}, and in addition, it is number squeezed.
\end{subsection}
\begin{subsection}{Beam splitter entanglement}\label{sec43}
The nonclassical nature of noncommutative coherent states are quite obvious from the analysis that we have carried out in Sec. \ref{sec41} and \ref{sec42}. Here, we would like to verify our results from a completely different approach, by using a quantum beam splitter. A quantum beam splitter is a device which produces entangled states at its output ports, when at least one of the input ports is fed with nonclassical states. The output states of the quantum mechanical version of beam splitter are realised by a unitary operator $\mathcal{B}$ acting on the input states: \cite{Campos_Saleh_Teich}
\begin{equation}
\vert\text{out}\rangle = \mathcal{B}\vert\text{in}\rangle =\text{e}^{\frac{\theta}{2}(a^\dagger b e^{i\varphi}-ab^\dagger e^{-i \varphi)}}\vert\text{in}\rangle ,\qquad \vert\text{in}\rangle = \vert\psi_1\rangle \otimes \vert\psi_2\rangle,
\end{equation}
where $a, a^\dagger$ and $b, b^\dagger$ are the sets of ladder operators acting on the input fields $\vert\psi_1\rangle$ and $\vert\psi_2\rangle$, respectively. $\theta\in [0,\pi]$ parametrises the transmittance and reflectance, and $\varphi$ denotes the phase difference between the reflected and transmitted fields. For further details on the device, we refer the readers to \cite{Campos_Saleh_Teich,Kim_Son_Buzek_Knight,Gerry_Knight_Book,Dey_Hussin}. For our purpose, we would like to consider the noncommutative coherent states \textcolor{beamer@PRD}{(}\ref{NcCoherentAlt}\textcolor{beamer@PRD}{)} as one of the inputs; say at $\vert\psi_1\rangle$, while a vacuum state $\vert 0\rangle$ at the other. The output states in this case are computed to 
\begin{equation}\label{NcCoherentInput}
\vert\text{out}\rangle = \mathcal{B}\big(\vert\alpha,f,\phi\rangle\otimes\vert 0\rangle\big)=\frac{1}{\mathcal{N}(\alpha,f)}\displaystyle\sum_{n=0}^{\infty}\frac{\mathcal{C}(\alpha,n)}{\sqrt{n!}f(n)!}\mathcal{B}\big(\vert n\rangle\otimes\vert 0\rangle\big).
\end{equation}
The effect of the beam splitter operator on a bipartite input state composed of a Fock state $\vert n\rangle$ and a vacuum state $\vert 0\rangle$, is well known \cite{Kim_Son_Buzek_Knight} 
\begin{equation}\label{FockInput}
\mathcal{B}\big(\vert n\rangle\otimes\vert 0\rangle\big)=\displaystyle\sum_{q=0}^{n} \begin{pmatrix}
n \\ q
\end{pmatrix}^{1/2} t^qr^{n-q}\big(\vert q\rangle\otimes\vert n-q\rangle\big),
\end{equation}
where $t=\cos(\theta/2)$ and $r=-e^{i\phi}\sin(\theta/2)$ are the transmission and reflection amplitudes, respectively. Substituting Eq. \textcolor{beamer@PRD}{(}\ref{FockInput}\textcolor{beamer@PRD}{)} in \textcolor{beamer@PRD}{(}\ref{NcCoherentInput}\textcolor{beamer@PRD}{)} and following the similar steps as in Ref. \cite{Dey_Hussin}, we compute the reduced density matrix of the output states, such that the linear entropy becomes
\begin{eqnarray}
S &=& 1-\frac{1}{\mathcal{N}^4(\alpha,f)} \displaystyle\sum_{q=0}^\infty\displaystyle\sum_{s=0}^\infty\displaystyle\sum_{m=0}^{\infty-\text{max}(q,s)} \displaystyle\sum_{n=0}^{\infty-\text{max}(q,s)} \vert t\vert^{2(q+s)}\vert r\vert^{2(m+n)} \notag \\ 
&& \times\frac{\mathcal{C}(\alpha,m+q)\mathcal{C}^\ast(\alpha,m+s)\mathcal{C}(\alpha,n+s)\mathcal{C}^\ast(\alpha,n+q)}{q!s!m!n!f(m+q)!f(m+s)!f(n+s)!f(n+q)!}~.\label{Entropy}
\end{eqnarray}
Assuming the input states of the beam splitter to be nonclassical, we expect the output states to be entangled and, hence, a finite amount of linear entropy must be created. The results demonstrated in Fig. \ref{fig5} confirm our claim and establish the nonclassical nature of the noncommutative coherent states. In contrast, when $f(n)=1$, which corresponds to the case of ordinary harmonic oscillator, the output states are not entangled and naturally we do obtain a null entropy. The most interesting fact is that when we enhance the noncommutativity by increasing the value of the parameter $\tau$, the entanglement rises accordingly and, therefore, becomes more and more nonclassical as shown in left panel of Fig. \ref{fig5}. A similar type of result was found in \cite{Dey_Hussin}, however, for the sake of completeness we recall it here along with our main observations.
\begin{figure}[h]
\centering   \includegraphics[width=8.2cm,height=6.0cm]{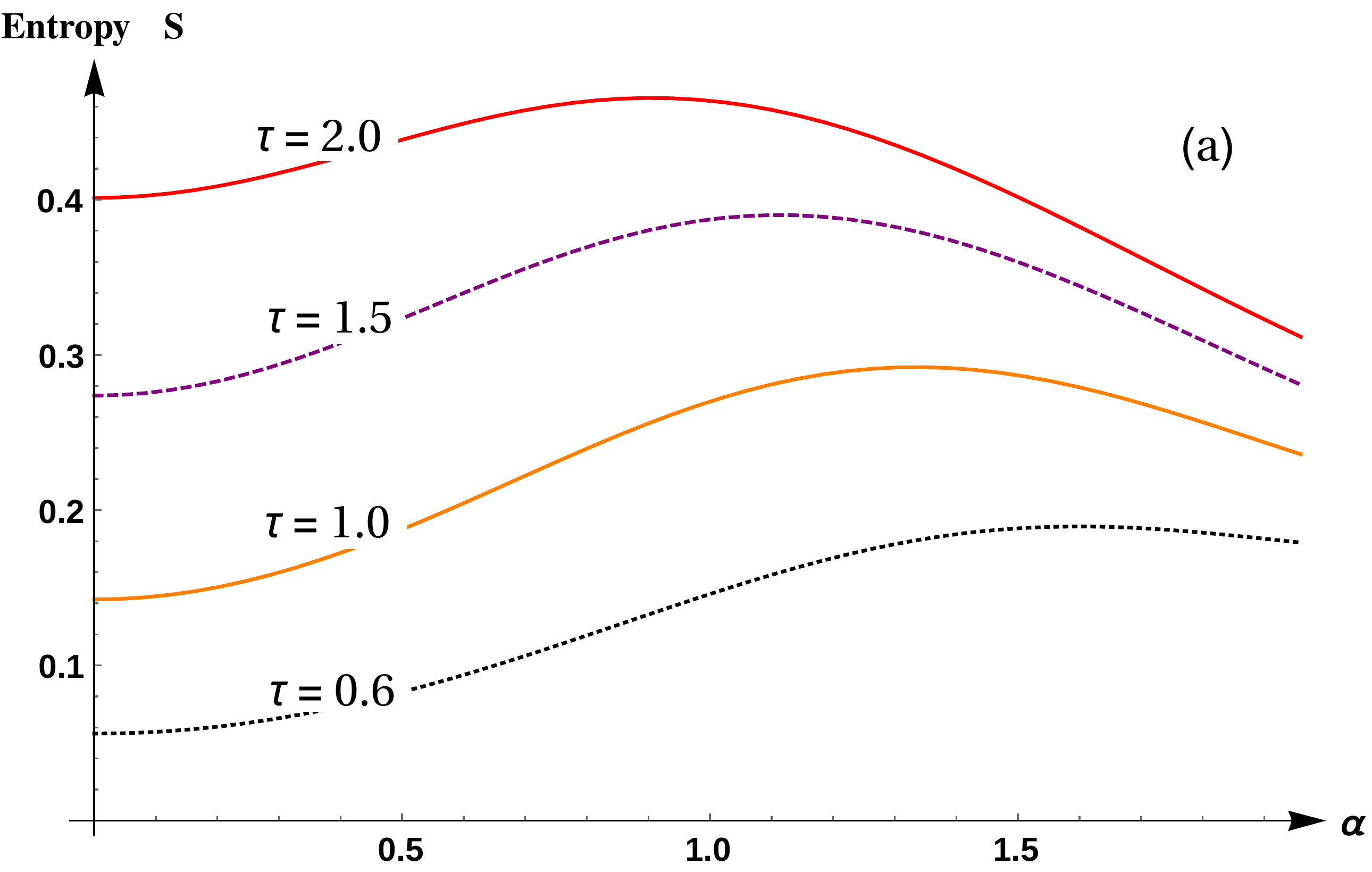}
\includegraphics[width=8.2cm,height=6.0cm]{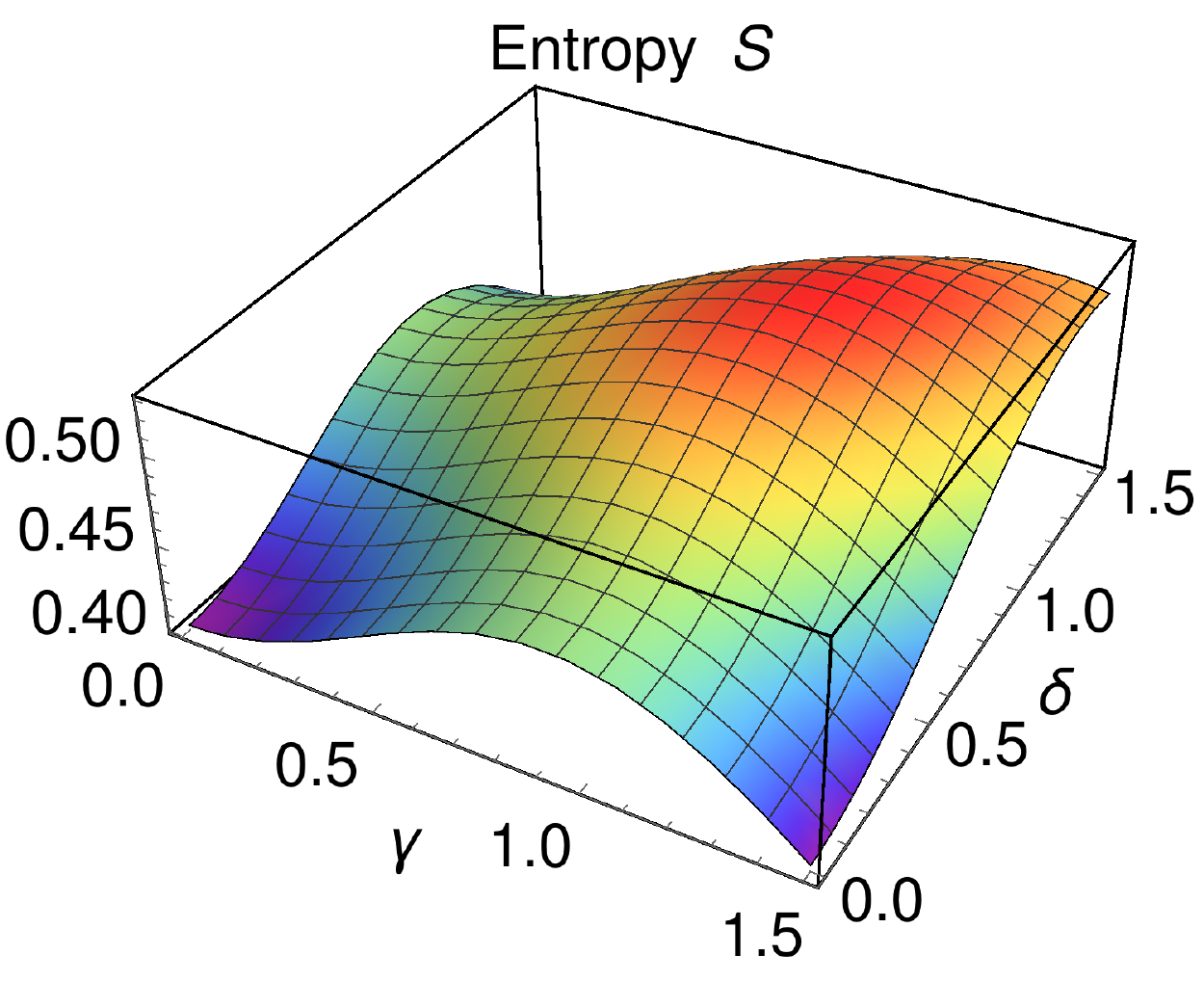}
\caption{\small{Linear entropy of noncommutative coherent states (a) for different values of $\tau$, (b) in the complex plane of $\alpha$, with $\alpha =\gamma+i\delta$ and $\tau =2$.}}
\label{fig5}
\end{figure}      
\end{subsection}
\end{section}  
\begin{section}{Nonclassicality of cat states and their superiorities}\label{sec5}
We now compute the expectation values of the ladder operators $a$ and $a^\dagger$ for the cat states \textcolor{beamer@PRD}{(}\ref{NcCat}\textcolor{beamer@PRD}{)}
\begin{alignat}{1}
&_\pm\langle\alpha,f,\phi\vert a\vert\alpha,f,\phi\rangle_\pm =~_\pm\langle\alpha,f,\phi\vert a^\dagger\vert\alpha,f,\phi\rangle_\pm = 0.
\end{alignat}
We consider the quadrature operators \textcolor{beamer@PRD}{(}\ref{quadrature}\textcolor{beamer@PRD}{)} once again and follow the similar steps as Sec. \ref{sec4} to compute the expectation values of the noncommutative quadrature operators 
\begin{alignat}{1} \label{YC}
_\pm\langle\alpha,f,\Phi\vert Y\vert\alpha,f,\Phi\rangle_{\eta,\pm} &=~_\pm\langle\alpha,f,\Phi\vert Z\vert\alpha,f,\Phi\rangle_{\eta,\pm} = 0.
\end{alignat} 
More challenging are the computations of the expectation values of the square of the quadrature operators. After carring out quite lengthy calculations, we arrive at  
\begin{equation}\label{Y2C}
_\pm\langle\alpha,f,\Phi\vert Y^2\vert\alpha,f,\Phi\rangle_{\eta,\pm} = \frac{M_1^+\pm M_1^-}{\hat{\mathcal{N}}},\quad _\pm\langle\alpha,f,\Phi\vert Z^2\vert\alpha,f,\Phi\rangle_{\eta,\pm} =\frac{M_2^+\pm M_2^-}{\hat{\mathcal{N}}},
\end{equation}
where
\begin{alignat}{1}
M_1^\pm &= \frac{e^{\pm\vert\alpha\vert^2}}{4}\Big[2\mu_\pm+\tau\big\{6-\mu_\mp-2(\alpha^2+\alpha^{\ast 2})^2-\lambda_\pm\big\}\Big],\notag \\
M_2^\pm &= \frac{e^{\pm\vert\alpha\vert^2}}{4}\Big[8-2\mu_\mp+\tau\big\{\mu_\mp-2+2(\alpha^2-\alpha^{\ast 2})^2+\lambda_\pm\mp 8\vert \alpha\vert^2-10\vert \alpha\vert^4\mp 4\vert \alpha\vert^6\big\}\Big],\notag
\end{alignat}
with $\mu_\pm=2+2(\alpha\pm\alpha^\ast)^2,~\lambda_\pm=\pm 4\vert \alpha\vert^2(\alpha^2+\alpha^{\ast 2})+\vert \alpha\vert^4(1+\alpha^2+\alpha^{\ast 2})\pm 2\vert \alpha\vert^6$ and $\hat{\mathcal{N}}=\mathcal{N}^2(\alpha,f)\mathcal{N}^2(\alpha,f)_\pm$.
\begin{subsection}{Even cat states}
Assembling \textcolor{beamer@PRD}{(}\ref{YC}\textcolor{beamer@PRD}{)} and \textcolor{beamer@PRD}{(}\ref{Y2C}\textcolor{beamer@PRD}{)}, we compute the square of the uncertainties for both of the quadratures for even cat states  
\begin{equation}\label{UncertEven}
\Delta Y^2_+ = R_++U_+, \quad \Delta Z^2_+ = R_+-\tilde{U}_+, \quad \Delta Y_+^2\Delta Z_+^2 = R_+^2+R_+(U_+-\tilde{U}_+)-U_+\tilde{U}_+,
\end{equation}
where
\begin{alignat}{1}
R_+ &= \frac{1}{2}\Big[1+\tau~_+\langle\alpha,f,\Phi\vert Z^2\vert\alpha,f,\Phi\rangle_{\eta,+}\Big] = \frac{1}{2}+\frac{\tau}{4}\Big[1-\alpha^2-\alpha^{\ast 2}+2\vert\alpha\vert^2\tanh(\vert\alpha\vert^2)\Big], \label{Eq5.5}\\
U_+ &= \frac{\alpha^2+\alpha^{\ast 2}}{2}+\vert\alpha\vert^2\tanh(\vert\alpha\vert^2)+\frac{\tau}{4}\Big[1-(\alpha^2-\alpha^{\ast 2})^2+2\vert\alpha\vert^2\tanh(\vert\alpha\vert^2)-\frac{4\vert\alpha\vert^4}{\cosh^2(\vert\alpha\vert^2)}\Big],\label{Eq5.6} \\
\tilde{U}_+ &= \frac{(\alpha-\alpha^\ast)^2(1-\tau)}{2}+\frac{\tau}{4}\Big[1+2\vert\alpha\vert^2-(\alpha^2-\alpha^{\ast 2})^2\Big]+\frac{\vert\alpha\vert^2(2-3\tau+4\tau\vert\alpha\vert^2)}{(1+e^{2\vert\alpha\vert^2})}-\frac{4\tau\vert\alpha\vert^4}{(1+e^{2\vert\alpha\vert^2})^2}.\label{Eq5.7}
\end{alignat}
$R_+$ in Eq. \textcolor{beamer@PRD}{(}\ref{Eq5.5}\textcolor{beamer@PRD}{)} denotes the right hand side of the generalised uncertainty relation \textcolor{beamer@PRD}{(}\ref{GUR}\textcolor{beamer@PRD}{)}. The uncertainty relation is valid if the following condition holds:
\begin{equation}\label{validity}
R_+(U_+-\tilde{U}_+)-U_+\tilde{U}_+\geq 0.
\end{equation}
The restriction is not a surprise for us, as our work is based on a perturbative model. However, the relation \textcolor{beamer@PRD}{(}\ref{validity}\textcolor{beamer@PRD}{)} is true for almost all values of $\alpha$ and $\tau$, for instance if $\text{Im}[\alpha]-\text{Re}[\alpha]\geq 0.1,~\forall \tau$ with $(\text{Re}[\alpha])_\text{min}=0.9$ and $(\text{Im}[\alpha])_\text{min}=1.0$. In some of the figures we might cross the limit, but one can easily exclude that particular small portion and; see, that the results hold in general. In fact, all of the results improve much more if one restricts the systems to the allowed region. Nevertheless, it is very clear from Eq. \textcolor{beamer@PRD}{(}\ref{UncertEven}\textcolor{beamer@PRD}{)}-\textcolor{beamer@PRD}{(}\ref{Eq5.7}\textcolor{beamer@PRD}{)} that the quadrature $Z$ is susceptible to squeezing in our case, while the quadrature $Y$ is not. What remains is to verify the positivity of $U_+$ and $\tilde{U}_+$. However, before doing that, let us quickly verify the compatibility of our findings in the limiting situation. In the limit; $\tau\rightarrow 0$, the Eq. \textcolor{beamer@PRD}{(}\ref{Eq5.6}\textcolor{beamer@PRD}{)} and \textcolor{beamer@PRD}{(}\ref{Eq5.7}\textcolor{beamer@PRD}{)} reduce to 
\begin{eqnarray}
\big(U_+\big)_{\text{ho}} &=& \frac{1}{2}\Big(\alpha^2+\alpha^{\ast 2}+2\vert\alpha\vert^2\tanh(\vert\alpha\vert^2)\Big), \label{Eq5.8}\\
\big(\tilde{U}_+\big)_{\text{ho}} &=& \frac{1}{2}\Big(\alpha^2+\alpha^{\ast 2}-2\vert\alpha\vert^2\tanh(\vert\alpha\vert^2)\Big),\label{Eq5.9}  
\end{eqnarray} 
which coincide precisely with the results of the even cat states of ordinary harmonic oscillator \cite{Gerry_Knight}.
\begin{figure}[h]
\centering   \includegraphics[width=8.2cm,height=6.0cm]{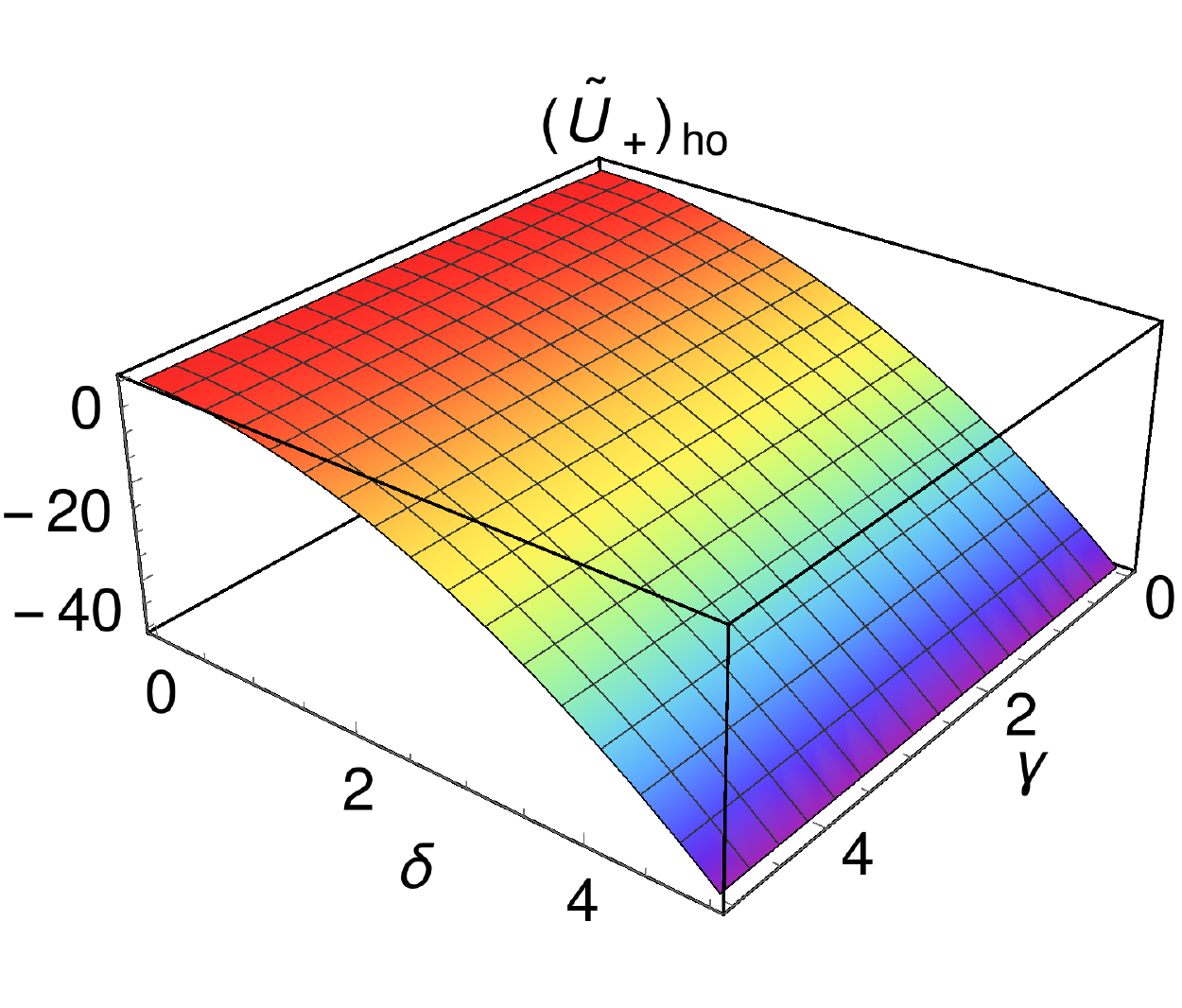}
\includegraphics[width=8.2cm,height=6.0cm]{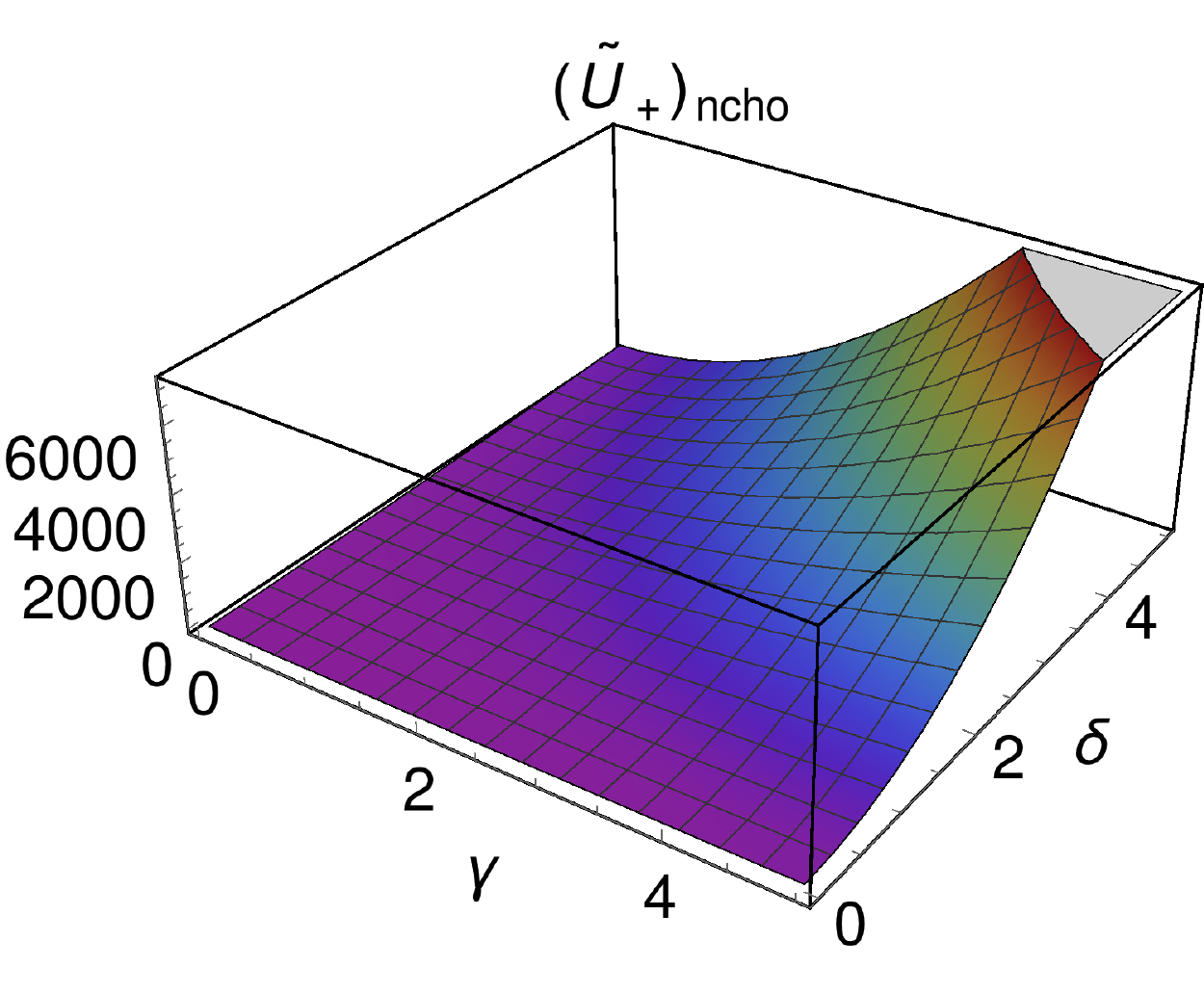}
\caption{\small{Expansion of quadrature $z$ for ordinary harmonic oscillator (left panel) versus squeezing of quadrature $Z$ for noncommutative harmonic oscillator for $\tau =5$ (right panel) in the complex plane of $\alpha$, with $\alpha =\gamma+i\delta$.}}
\label{fig1}
\end{figure}
Numerically, we check that $U_+$ \textcolor{beamer@PRD}{(}\ref{Eq5.6}\textcolor{beamer@PRD}{)} and $(U_+)_{\text{ho}}$ \textcolor{beamer@PRD}{(}\ref{Eq5.8}\textcolor{beamer@PRD}{)} are always positive. However, $(\tilde{U}_+)_\text{ho}$ in Eq. \textcolor{beamer@PRD}{(}\ref{Eq5.9}\textcolor{beamer@PRD}{)} becomes negative, when one considers $\alpha$ to be complex, which indicates that in general there is no quadrature squeezing for the even cat states for ordinary harmonic oscillator. Although many authors have claimed the existence of quadrature squeezing, however, for special cases; e.g. considering $\alpha$ as real \cite{Gerry_Knight}. For convenience, we have depicted the general results for the ordinary harmonic oscillator in the complex plane in the left panel of Fig. \ref{fig1}. Quite surprisingly, the results change drastically in noncommutative spaces and one notices that $\tilde{U}_+$ \textcolor{beamer@PRD}{(}\ref{Eq5.7}\textcolor{beamer@PRD}{)} becomes positive everywhere, as shown in the right panel of Fig. \ref{fig1}. Thus, the quadrature $Z$ is squeezed for the noncommutative case, while this is not always true for the usual space. 

Let us now study the other nonclassical properties of the even cat states. To verify whether the number squeezing exists, we compute
\begin{eqnarray}
_+\langle N\rangle_+ &=& (1-\tau)\vert\alpha\vert^2\tanh(\vert\alpha\vert^2)+\tau \vert\alpha\vert^4 \Big(\tanh^2(\vert\alpha\vert^2)-\frac{3}{2}\Big), \\
_+\langle N^2\rangle_+&=& \vert\alpha\vert^4+(1-\tau-\tau\vert\alpha\vert^4)\vert\alpha\vert^2\tanh(\vert\alpha\vert^2)+\tau\vert\alpha\vert^4 \big(\tanh^2(\vert\alpha\vert^2)-4\big),
\end{eqnarray}
such that the Mandel parameter turns out to be
\begin{equation}\label{MandelEven}
Q_+ =\frac{\vert\alpha\vert^2}{2\sinh(2\vert\alpha\vert^2)}\Big[4-5\tau-\tau\cosh(2\vert\alpha\vert^2)\Big]+\frac{\tau \vert\alpha\vert^4}{\sinh^2(2\vert\alpha\vert^2)}\Big[1+5\cosh(2\vert\alpha\vert^2)\Big].
\end{equation}
The right panel of Fig. \ref{fig3} shows a plot of the Mandel parameter \textcolor{beamer@PRD}{(}\ref{MandelEven}\textcolor{beamer@PRD}{)} in the complex plane of $\alpha$, for a fixed value of the noncommutative parameter $\tau $. 
\begin{figure}
\centering   \includegraphics[width=8.2cm,height=6.0cm]{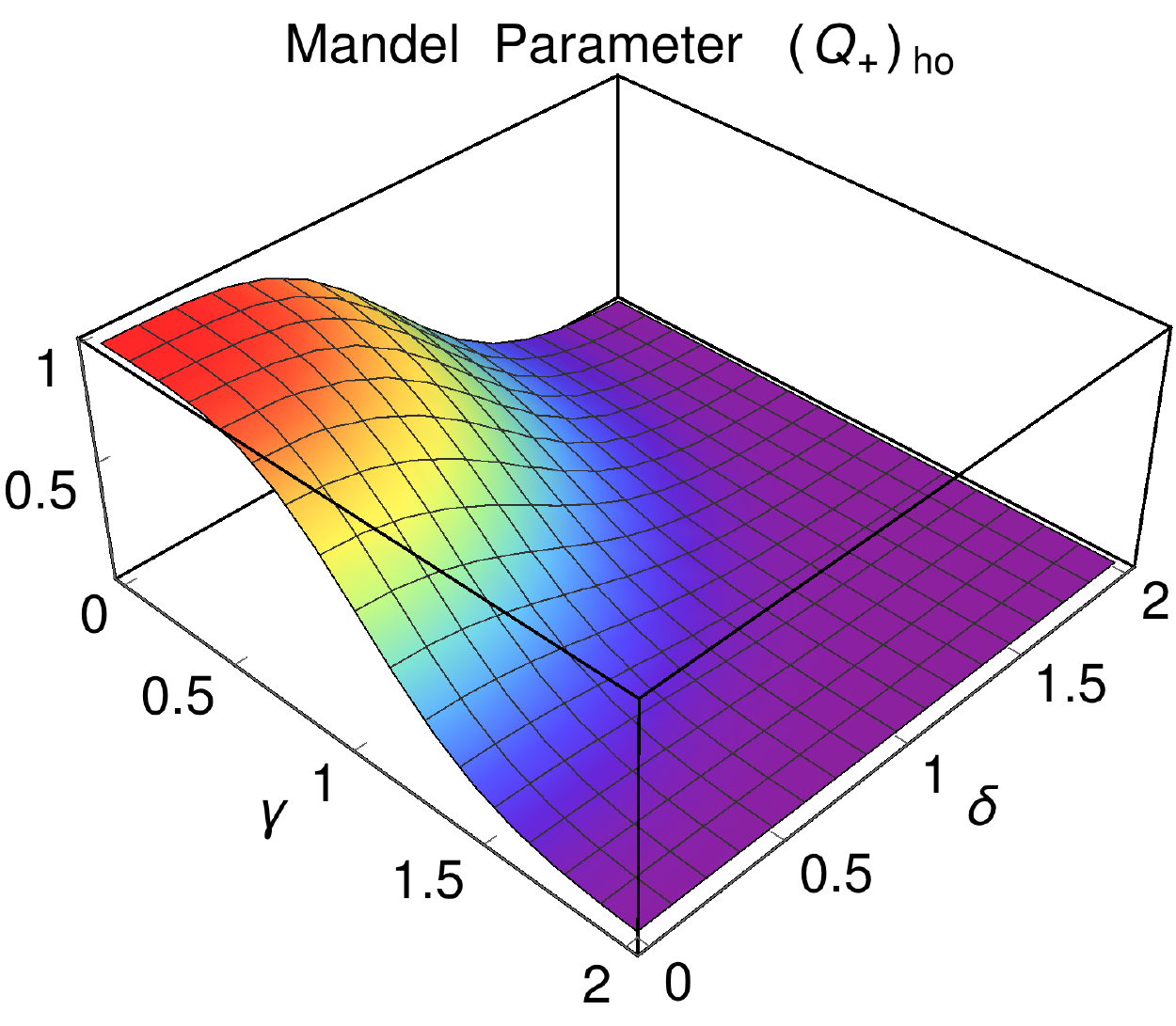}
\includegraphics[width=8.2cm,height=6.0cm]{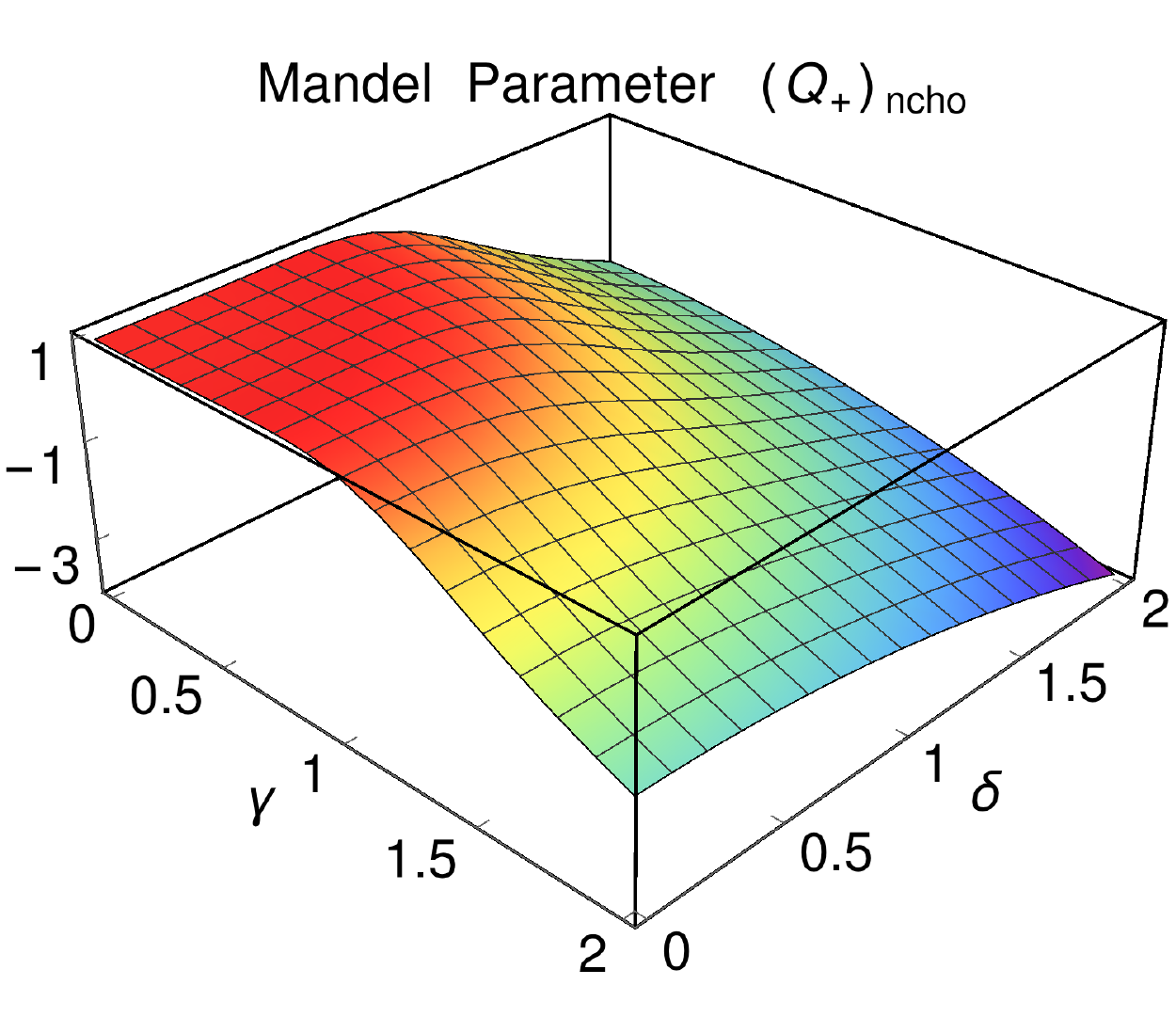}
\caption{\small{Mandel parameter of even cat states in complex plane of $\alpha$, with $\alpha =\gamma+i\delta$ for harmonic oscillator (left panel), for noncommutative harmonic oscillator for $\tau =1$ (right panel).}}
\label{fig3}
\end{figure}
Apart from very small values of $\alpha$, the Mandel parameter $Q_+$ is negative everywhere and, therefore, the number squeezing exists in some region of space. While, it is well known that for the even cat states of ordinary harmonic oscillator, the Mandel parameter $(Q_+)_{\text{ho}}=2\vert\alpha\vert^2 \text{csch}(2\vert\alpha\vert^2)>0~\forall~\alpha$ and, therefore, they are not number squeezed \cite{Gerry_Knight_Book}; see left panel of Fig. \ref{fig3}. It clearly indicates that the even cat states in noncommutative space are much more nonclassical than that of the usual space. Similar type of results were found by one of the authors recently \cite{Dey}, however, for $q$-deformed noncommutative space. A mutual comparison of photon distributions of even cat states, $P_{+}(n):=\vert_+\langle n\vert\alpha,f,\phi\rangle_+\vert^2$ and the coherent states, $P(n):=\vert\langle n\vert\alpha,f,\phi\rangle\vert^2$ in noncommutative space are depicted in Fig. \ref{fig8}\textcolor{beamer@PRD}{(a)}. The highly oscillating photon distribution for the cat states provides a strong indication of the nonclassicality of the states \cite{Gerry_Knight_Book}. 

Therefore, it seems that the even cat states are slightly more nonclassical than the coherent states. Let us quickly verify our claim by computing the entanglement. We follow the similar procedure as analysed in Sec. \ref{sec43}, but insert the even cat states instead of coherent states at one of the inputs. While, the other input remains same, i.e. a vacuum state $\vert 0\rangle$. The outcome is shown in the right panel of Fig. \ref{fig6}. One can compare the entanglement of even cat states with that of the coherent states from Fig. \ref{fig5} to observe the difference between these two. Nevertheless, the left panel of Fig. \ref{fig6} is more interesting, where the entanglement of noncommutative even cat states is compared with that of the ordinary cat states. The mutual comparison clearly indicates the superiorities of utilising noncommutative models instead of usual quantum mechanical models.   
\begin{figure}
\centering   \includegraphics[width=8.2cm,height=6.0cm]{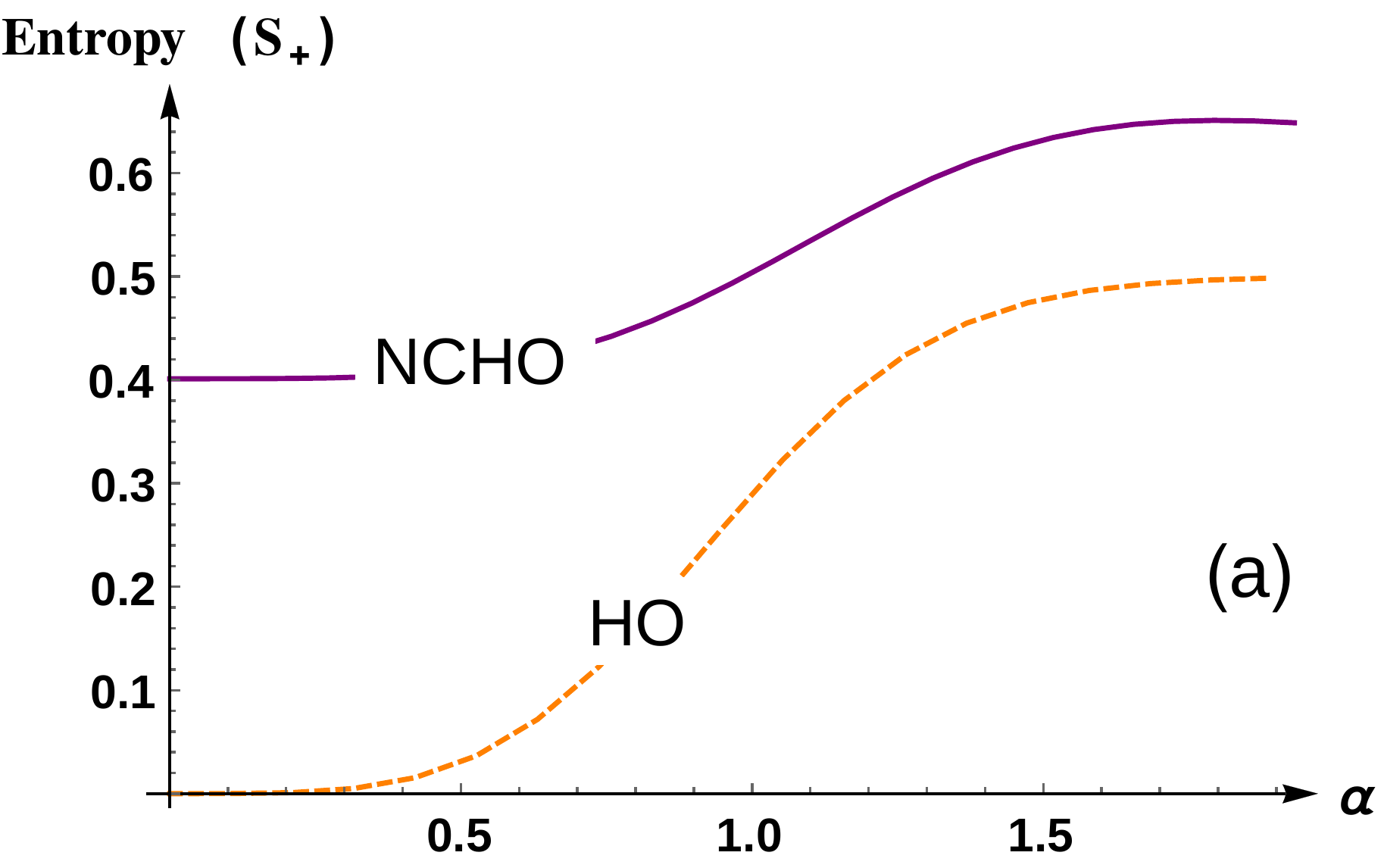}
\includegraphics[width=8.2cm,height=6.0cm]{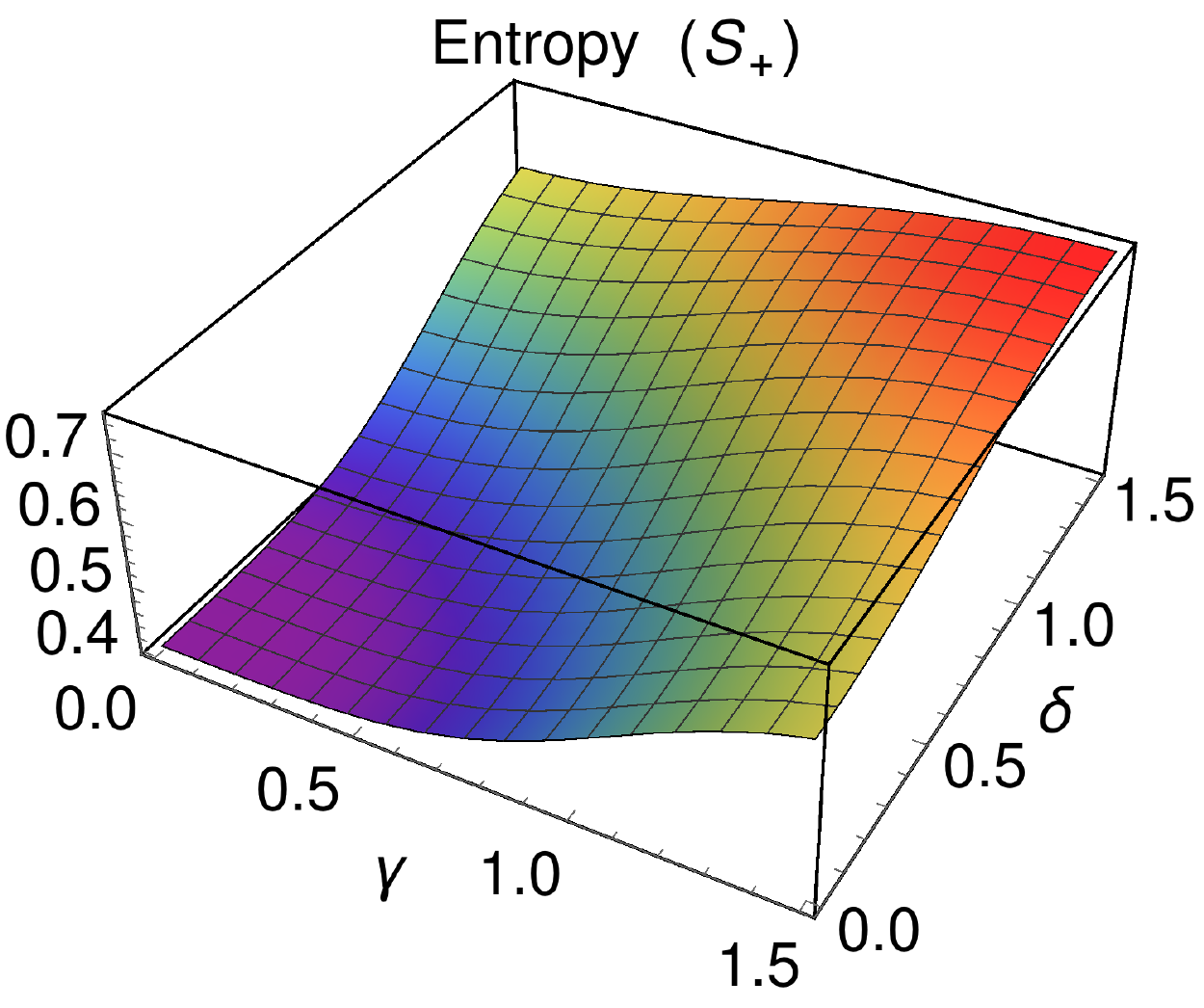}
\caption{\small{Linear entropy for even cat states (a) of ordinary harmonic oscillator (orange, dashed) versus noncommutative harmonic oscillator (purple, solid), (b) of noncommutative harmonic oscillator in the complex plane of $\alpha$, with $\alpha =\gamma+i\delta$ and $\tau =2$.}}
\label{fig6}
\end{figure} 
\end{subsection}
\begin{subsection}{Odd cat states}
Next, we assemble \textcolor{beamer@PRD}{(}\ref{YC}\textcolor{beamer@PRD}{)}-\textcolor{beamer@PRD}{(}\ref{Y2C}\textcolor{beamer@PRD}{)} to compute the square of the uncertainties for the quadratures for odd cat states
\begin{equation}
\Delta Y^2_- = R_-+U_-, \quad \Delta Z^2 = R_--\tilde{U}_-, \quad \Delta Y_-^2\Delta Z_-^2 = R_-^2+R_-(U_--\tilde{U}_-)-U_-\tilde{U}_-,
\end{equation}
with
\begin{alignat}{1}
R_- &= \frac{1}{2}\Big[1+\tau~_-\langle\alpha,f,\Phi\vert Z^2\vert\alpha,f,\Phi\rangle_{\eta,-}\Big] = \frac{1}{2}+\frac{\tau}{4}\Big[1-\alpha^2-\alpha^{\ast 2}+2\vert\alpha\vert^2\coth(\vert\alpha\vert^2)\Big], \label{Eq514} \\
U_- &= \frac{\alpha^2+\alpha^{\ast 2}}{2}+\frac{\vert\alpha\vert^2}{\tanh(\vert\alpha\vert^2)}+\frac{\tau}{4}\Big[1-(\alpha^2-\alpha^{\ast 2})^2+\frac{2\vert\alpha\vert^2}{\tanh(\vert\alpha\vert^2)}+\frac{4\vert\alpha\vert^4}{\sinh^2(\vert\alpha\vert^2)}\Big], \\
\tilde{U}_- &= \frac{\alpha^2+\alpha^{\ast 2}}{2}-\frac{\vert\alpha\vert^2}{\tanh(\vert\alpha\vert^2)}+\frac{\tau}{4}\Big[1-\alpha^2-\alpha^{\ast 2}-(\alpha^2-\alpha^{\ast 2})^2+\frac{6\vert\alpha\vert^2}{\tanh(\vert\alpha\vert^2)}-\frac{4\vert\alpha\vert^4}{\sinh^2(\vert\alpha\vert^2)}\Big].
\end{alignat}
$R_-$ in Eq. \textcolor{beamer@PRD}{(}\ref{Eq514}\textcolor{beamer@PRD}{)} represents the right hand side of the generalised uncertainty relation \textcolor{beamer@PRD}{(}\ref{GUR}\textcolor{beamer@PRD}{)}.
\begin{figure}
\centering   \includegraphics[width=8.2cm,height=6.0cm]{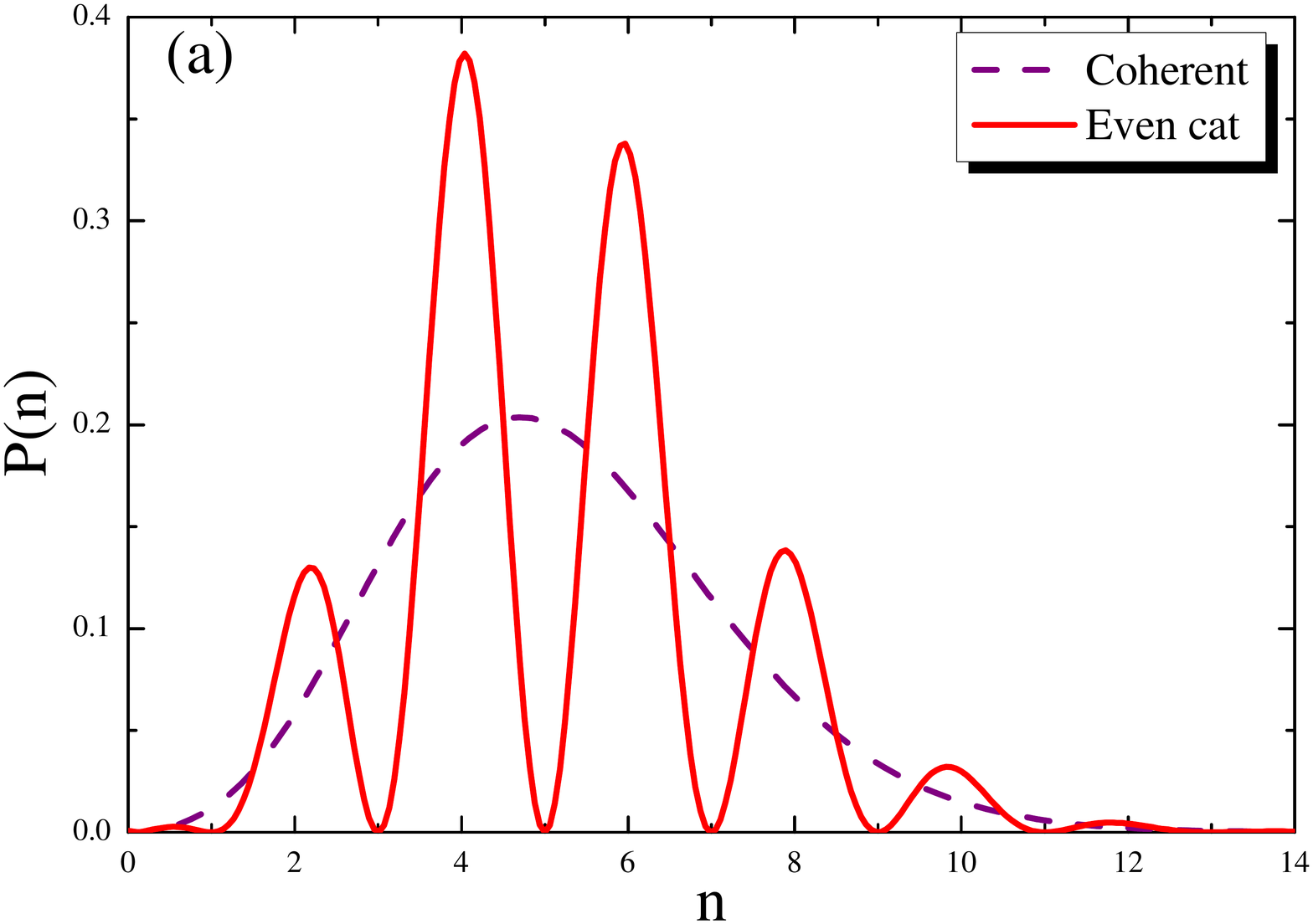}
\includegraphics[width=8.2cm,height=6.0cm]{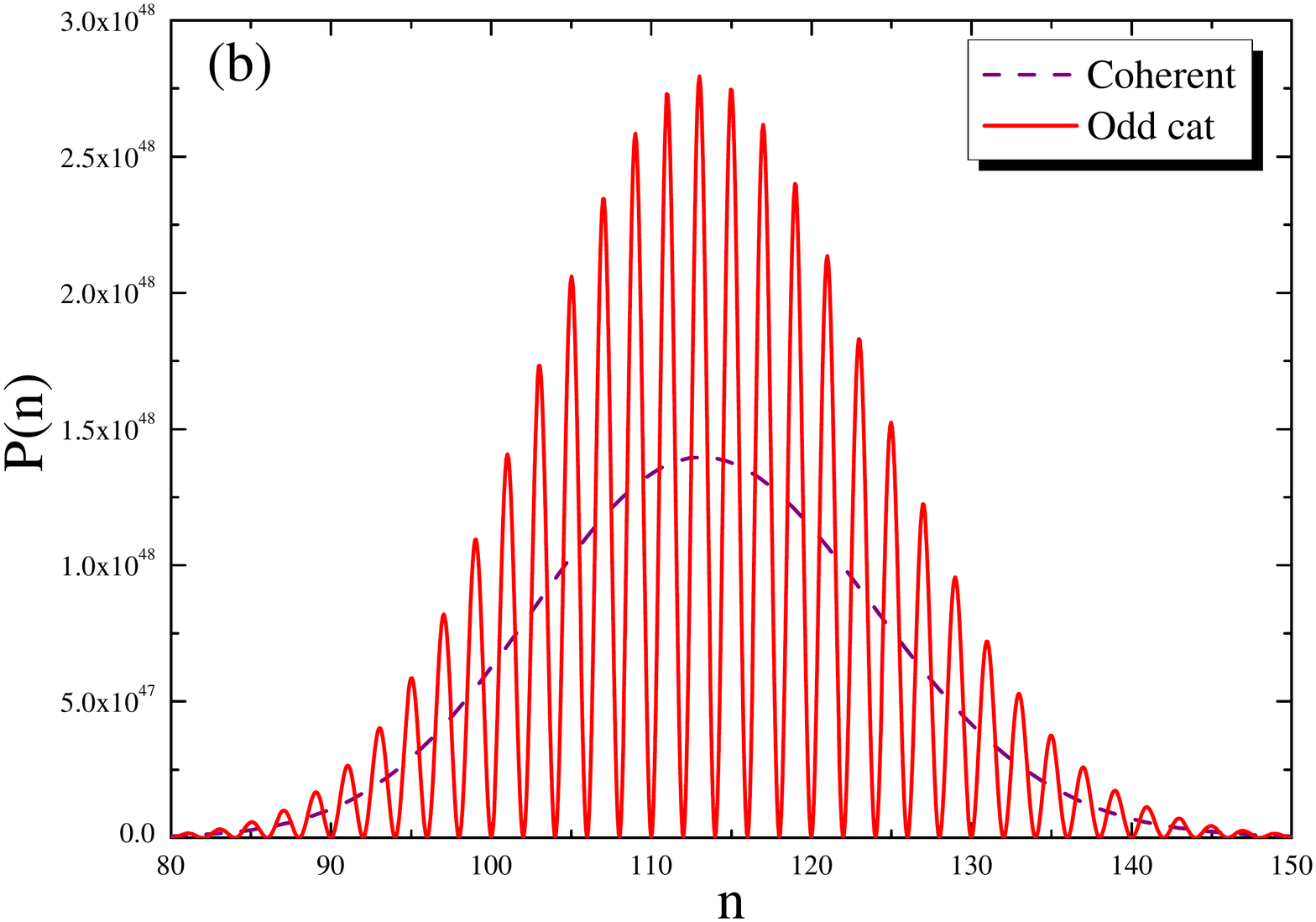}
\caption{\small{Photon distribution functions of coherent states (purple, dashed) versus (a) even cat states (red, solid) for $\alpha=1.2+1.5 i$, (b) odd cat states (red, solid) for $\alpha=1.2+10.5 i$ in noncommutative space for $\tau=10$.}}
\label{fig8}
\end{figure}
A similar analysis to the even cat states shows that the quadrature squeezing exists for the odd case as well.
\begin{figure}
\centering   \includegraphics[width=8.2cm,height=6.0cm]{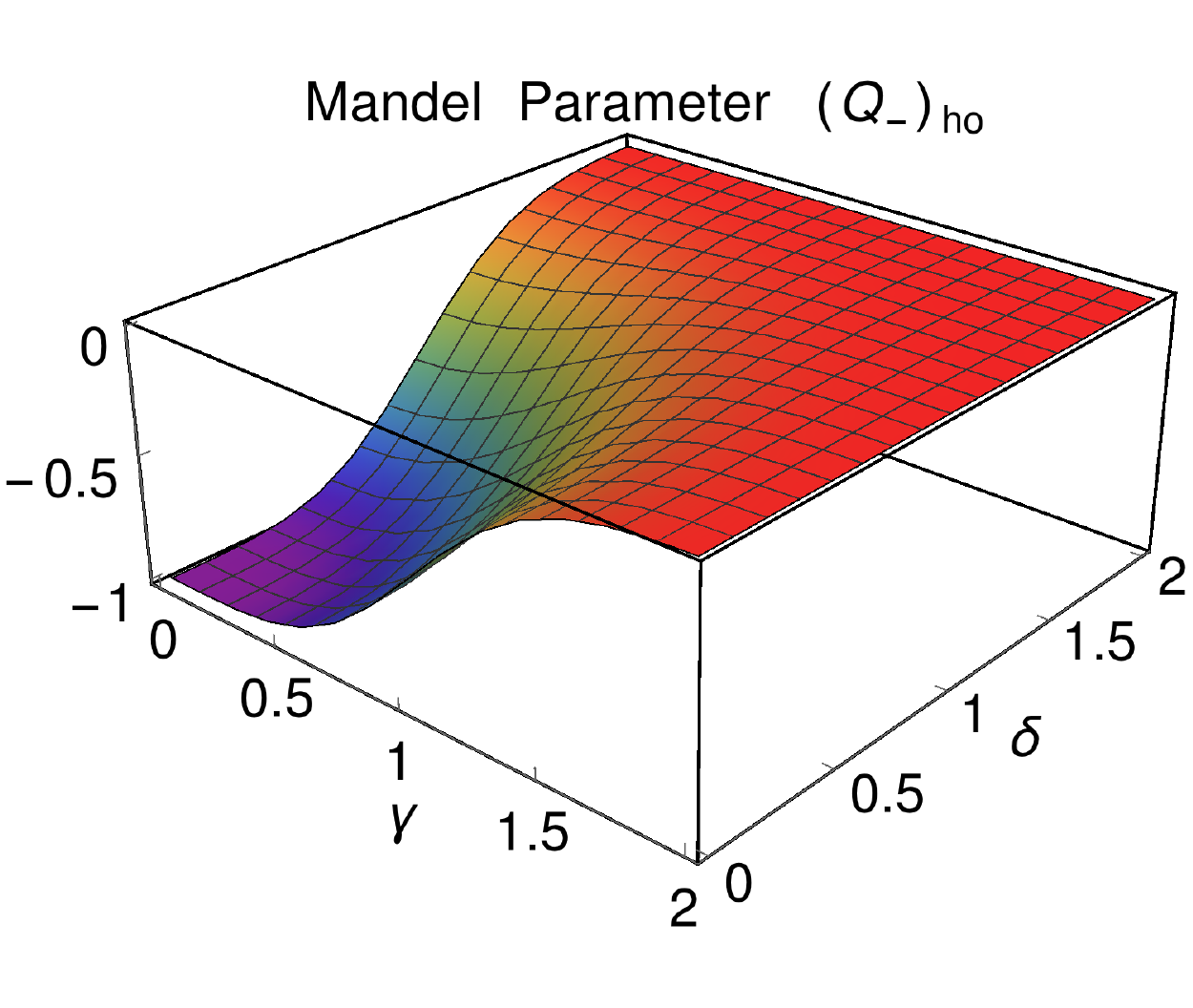}
\includegraphics[width=8.2cm,height=6.0cm]{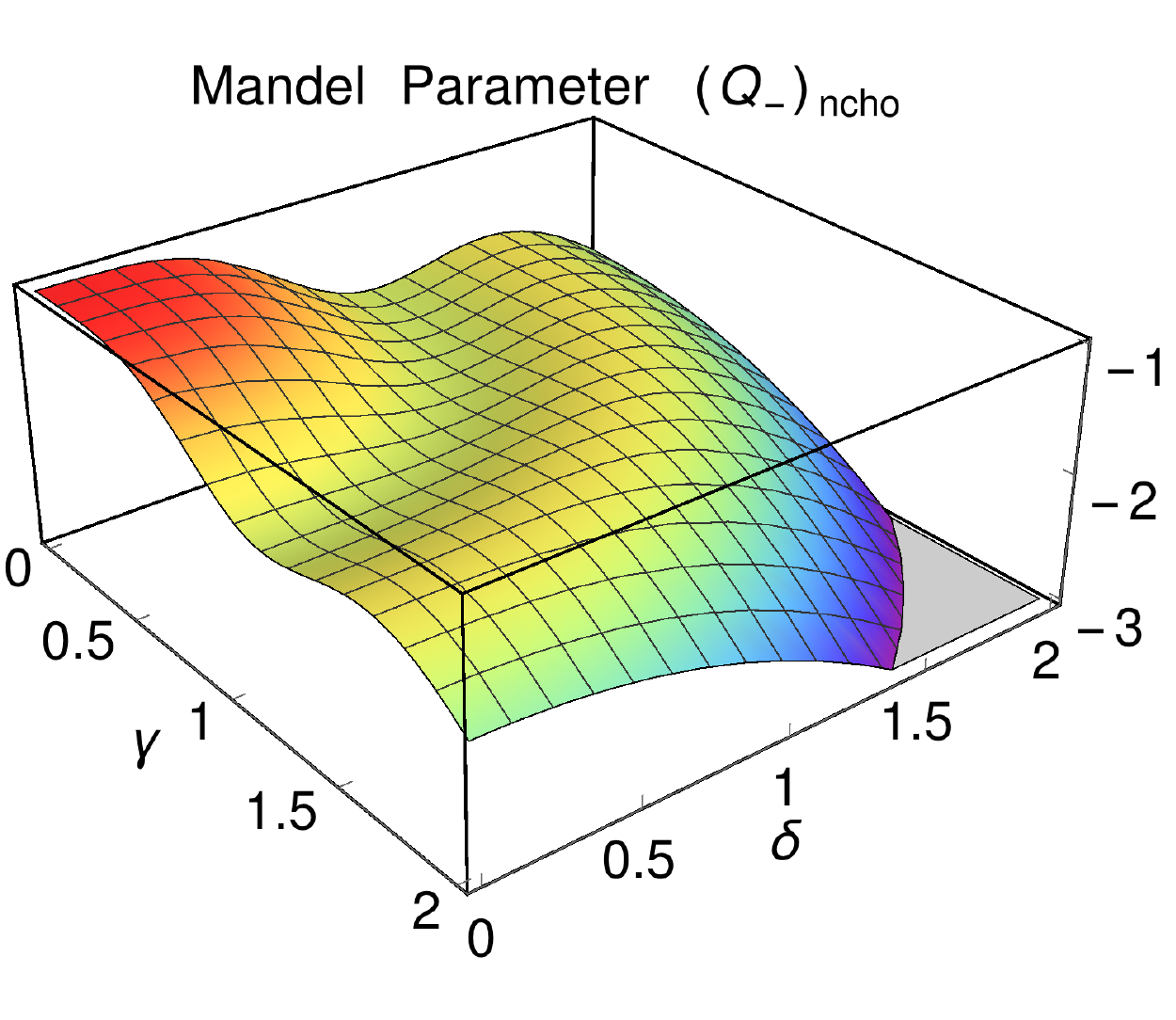}
\caption{\small{Mandel parameter of odd cat states in complex plane of $\alpha$, with $\alpha =\gamma+i\delta$ for harmonic oscillator (left panel), for noncommutative harmonic oscillator for $\tau =1$ (right panel).}}
\label{fig4}
\end{figure}
That means, $U_-$ and $\tilde{U}_-$ are always positive showing the similar nature as $U_+$ and $\tilde{U}_+$, respectively, which we do not present here. It is easy to check that for the case of ordinary harmonic oscillator, i.e. in the limiting condition, $\tau\rightarrow 0$, $U_-$ remains positive. However, $\tilde{U}_-$ becomes negative, for any values of $\alpha$, irrespective of whether it is real or complex, suggesting no quadrature squeezing. More exciting finding for the noncommutative odd cat states is that because of the photon antibunching, the Mandel parameter 
\begin{equation}
Q_-=-\frac{\vert\alpha\vert^2}{2}\Big[\tau \tanh(\vert\alpha\vert^2)+\frac{4(1-\tau)}{\sinh(2\vert\alpha\vert^2)}+\frac{\tau\vert\alpha\vert^2}{\sinh^2(\vert\alpha\vert^2)}\Big(2+3\tanh^2(\vert\alpha\vert^2)\Big)\Big],
\end{equation}
becomes negative in the whole region of space, which means in this case the number squeezing exists everywhere; see right panel of Fig. \ref{fig4}. Although for the case of harmonic oscillator we also obtain squeezing in photon number distribution, $(Q_-)_{\text{ho}}=-2\vert\alpha\vert^2\text{csch}(2\vert\alpha\vert^2)$, as shown in the left panel of Fig. \ref{fig4}. However, the squeezing is much more improved in the noncommutative case. Moreover, the odd cat states of harmonic oscillator is not quadrature squeezed. Whereas, for noncommutative case, the two fold squeezing; i.e. the simultaneous quadrature squeezing and the number squeezing makes them more nonclassical. As per our expectation, the photon distribution is also oscillating, which is shown in Fig. \ref{fig8}\textcolor{beamer@PRD}{(b)}. In fact, the odd cat states provide all the properties that a perfect nonclassical states should own; and therefore, they are the best nonclassical states among all the states that we have discussed here. Once again, the analysis can be verified by computing the quantum entanglement; see right panel of Fig. \ref{fig7}. According to our anticipation, we obtain the highest amount of linear entropy among all other cases that we have analysed in noncommutative space. On the other hand, odd cat states in noncommutative space provide better entanglement than the odd cat states of the ordinary harmonic oscillator as usual, which is no longer a surprise for us; see, left panel of Fig. \ref{fig7}.
\begin{figure}
\centering   \includegraphics[width=8.2cm,height=6.0cm]{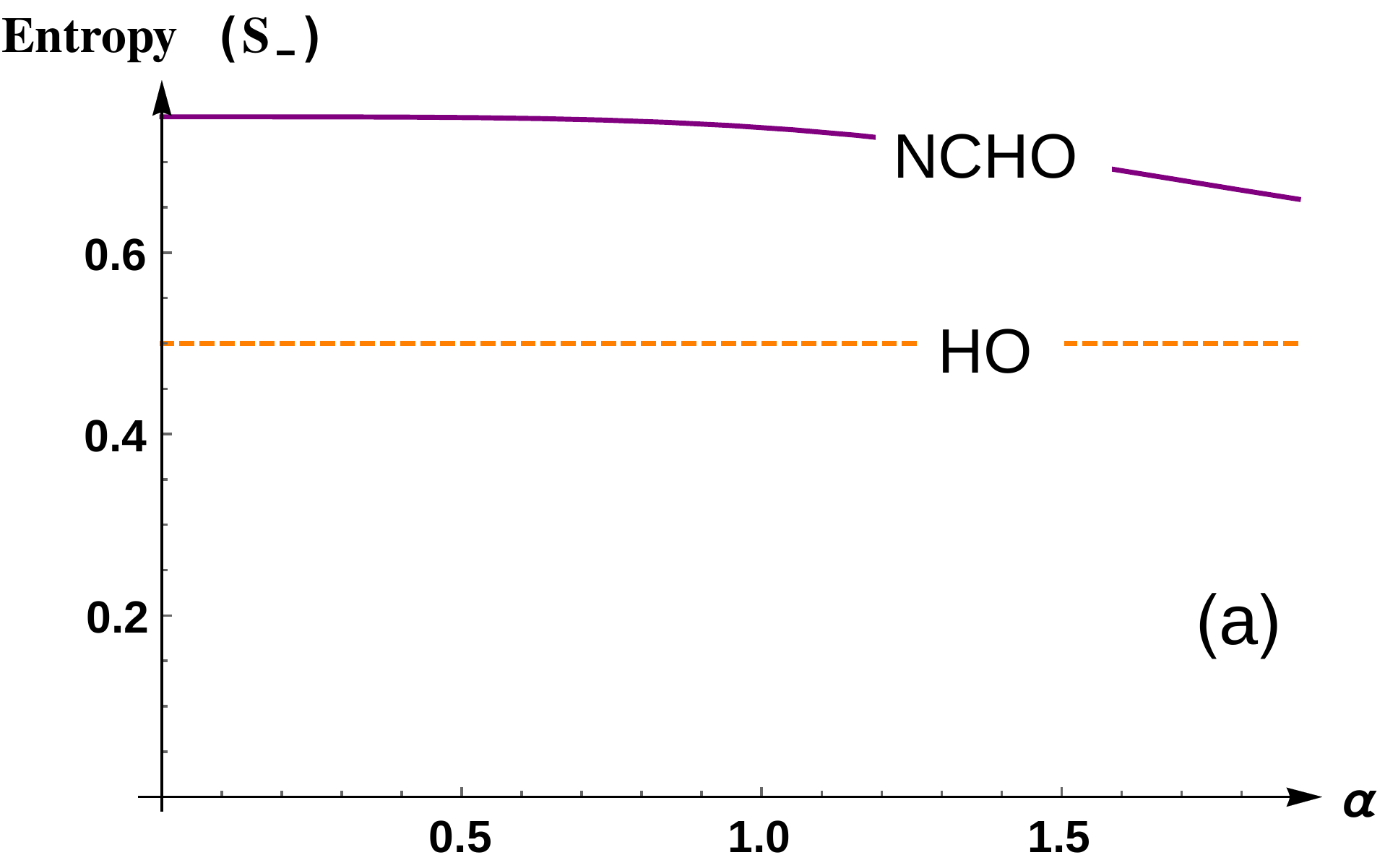}
\includegraphics[width=8.2cm,height=6.0cm]{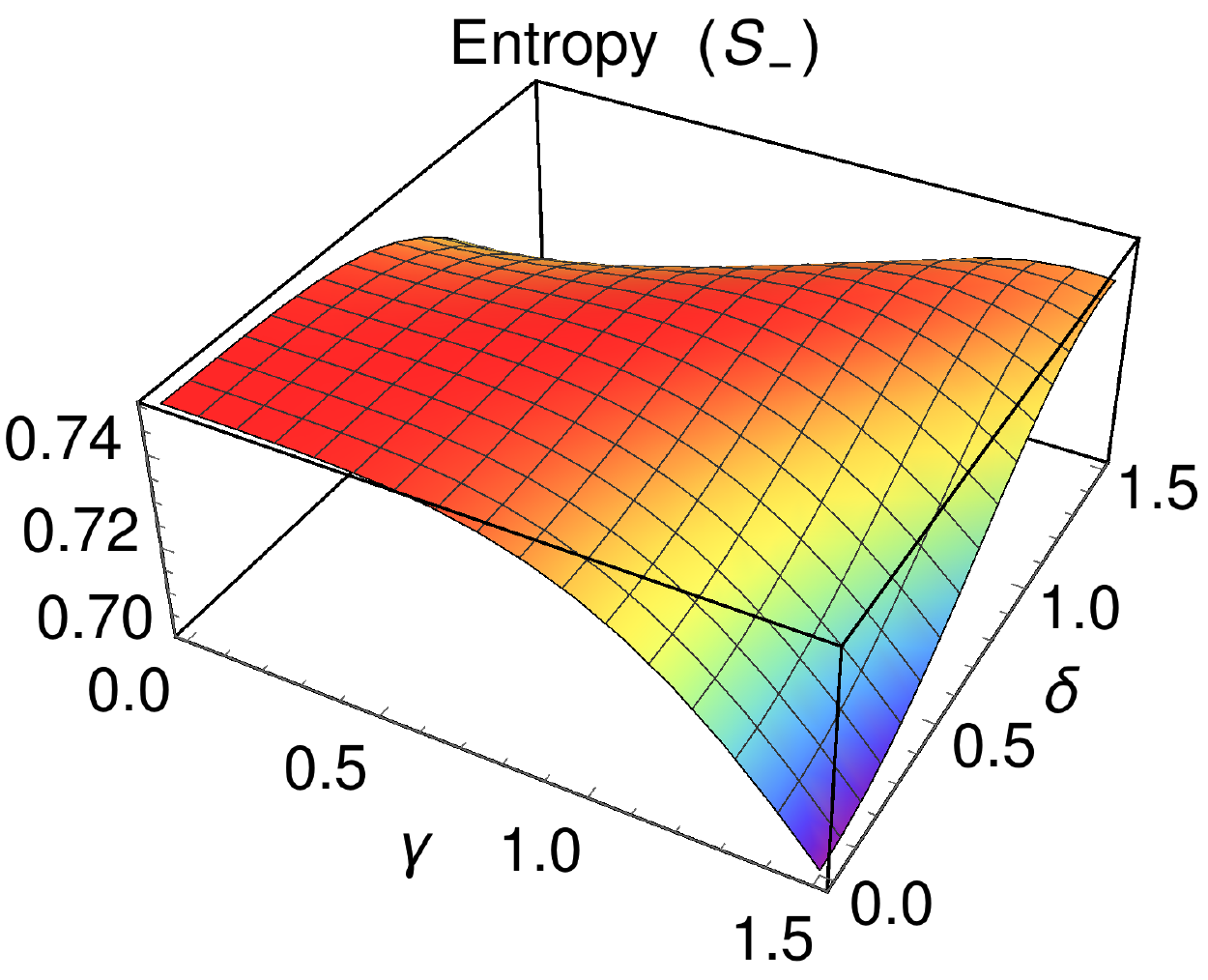}
\caption{\small{Linear entropy for odd cat states (a) of ordinary harmonic oscillator (orange, dashed) versus noncommutative harmonic oscillator (purple, solid), (b) of noncommutative harmonic oscillator in the complex plane of $\alpha$, with $\alpha =\gamma+i\delta$ and $\tau =2$.}}
\label{fig7}
\end{figure}
\end{subsection}
\end{section}      
\begin{section}{Conclusions}\label{sec6}
We have explored a convenient method of investigating the nonclassicality of states in a noncommutative structure resulting from the generalised uncertainty relation in a two-fold way. First, by analysing several nonclassical properties in a complete analytical manner; such as, quadrature squeezing, photon number squeezing and photon distribution function. Second, by computing the entanglement of the corresponding states by utilising a quantum mechanical beam-splitter. Both of the approaches lead us to the conclusion that the odd Schr\"odinger cat states are the best nonclassical states among all other states that we have discussed here, followed by the even cat states and coherent states, respectively. However, our principal observation is to identify the advantages of utilising noncommutative space-time structure instead of the usual quantum mechanical setting in the context of nonclassicality and entanglement. As for example, even and odd cat states in noncommutative space are found to be more nonclassical and entangled than that of the harmonic oscillator cat states. Coherent state in noncommutative space shows nonclassical behaviour,  while in contrast, the Glauber coherent states; i.e. the coherent states for ordinary harmonic oscillator do not possess these kind of properties. Moreover, the squeezing properties of noncommutative coherent states are not very common, they are minimum uncertainty quadrature and number squeezed states; i.e., the ideal squeezed states, which are not only very rare, but also to our knowledge, are the first finding in the literature. 

There are various ways in which one can extend our investigations. First, one can study many other nonclassical properties of our systems; such as, higher order squeezing \cite{Hong_Mandel}, Wigner function \cite{Wigner} etc. Second, it will be worth choosing some other type of noncommutative system and perform a similar analysis as presented here.

\vspace{0.5cm} \noindent \textbf{\large{Acknowledgements:}} S. D. is supported by the Postdoctoral Fellowship jointly funded by the Laboratory of Mathematical Physics of the Centre de Recherches Math{\'e}matiques (CRM) and by Prof. Syed Twareque Ali, Prof. Marco Bertola and Prof. V{\'e}ronique Hussin.

\end{section}


\end{document}